\documentclass[10pt,conference]{IEEEtran}
\pdfoutput=1
\usepackage{fancyhdr}
\usepackage{colortbl}
\usepackage[english]{babel}
\usepackage[utf8x]{inputenc}
\usepackage[T1]{fontenc}
\usepackage{courier}
\usepackage{listings}
\usepackage{color}
\usepackage{booktabs}
\usepackage{amssymb}
\usepackage{amsmath}
\usepackage{pifont}
\usepackage[numbers,sort]{natbib}
\usepackage{multirow}
\usepackage{multicol}
\usepackage[hyphens]{url}
\usepackage[breaklinks=true]{hyperref} 
\usepackage{adjustbox}
\usepackage{tabularx}
\usepackage{setspace}
\usepackage{xspace}
\usepackage{etoolbox}
\usepackage{changepage}
\usepackage{relsize}
\usepackage{arydshln} 
\usepackage{courier}
\usepackage{changepage}
\usepackage{ifthen}
\usepackage[sc]{caption}
\usepackage{tikz}

\DeclareRobustCommand\circledw[1]{\tikz[baseline=(char.base)]{\node[shape=circle,draw,inner sep=1pt ] (char) {\footnotesize \sffamily #1};}}

\newcommand{\cmark}{\ding{51}}
\newcommand{\xmark}{\ding{55}}

\definecolor{dkgreen}{rgb}{0,0.6,0}
\definecolor{gray}{rgb}{0.5,0.5,0.5}
\definecolor{mauve}{rgb}{0.58,0,0.82}

\makeatletter

\patchcmd{\@makecaption}{\scshape}{}{}{}

\makeatother

\begin{document}

\author{
\IEEEauthorblockN{
Ghada Dessouky$^\dagger$,
David Gens$^\dagger$,
Patrick Haney$^\ast$,
Garrett Persyn$^\ast$,
Arun Kanuparthi$^\circ$,\\
Hareesh Khattri$^\circ$,
Jason M. Fung$^\circ$,
Ahmad-Reza Sadeghi$^\dagger$,
Jeyavijayan Rajendran$^\ast$
}\\
\IEEEauthorblockA{
$^\dagger$Technische Universit\"at Darmstadt, Germany.\\
{\tt\{first.last@trust.tu-darmstadt.de\}}\\
$^\ast$Texas A\&M University, College Station, USA.\\
{\tt\{first.last@tamu.edu\}}\\
$^\circ$Intel Corporation, Hillsboro, OR. USA.\\
{\tt\{first.last@intel.com\}}\\
}}

\title{\bf When a Patch is Not Enough --- \\
HardFails: Software-Exploitable Hardware Bugs}
\maketitle

\begin{abstract}
Modern computer systems are becoming faster, more efficient, and increasingly interconnected with each generation. Consequently, these platforms also grow more complex, with continuously new features introducing the possibility of new bugs.
Hence, the semiconductor industry employs a combination of different verification techniques to ensure the security of System-on-Chip (SoC) designs during the development life cycle.
However, a growing number of increasingly sophisticated attacks are starting to leverage \emph{cross-layer bugs} by exploiting subtle interactions between hardware and software, as recently demonstrated through a series of real-world exploits with significant security impact that affected all major hardware vendors.

In this paper, we take a deep dive into microarchitectural security from a hardware designer's perspective by reviewing the existing approaches to detect hardware vulnerabilities during the design phase.
We show that a protection gap currently exists in practice that leaves chip designs vulnerable to software-based attacks.
In particular, existing verification approaches fail to detect specific classes of vulnerabilities, which we call \emph{HardFails}: these bugs evade detection by current verification techniques while being exploitable from software.
We demonstrate such vulnerabilities in real-world SoCs using RISC-V to showcase and analyze concrete instantiations of HardFails.
Patching these hardware bugs may not always be possible and can potentially result in a product recall.
We base our findings on two extensive case studies: the recent Hack@DAC~2018 hardware security competition, where 54 independent teams of researchers competed world-wide over a period of 12 weeks to catch inserted security bugs in SoC RTL designs, and an in-depth systematic evaluation of state-of-the-art verification approaches.
Our findings indicate that even combinations of techniques will miss high-impact bugs due to the large number of modules with complex interdependencies and fundamental limitations of current detection approaches.
We also craft a real-world software attack that exploits one of the RTL bugs from Hack@DAC that evaded detection and discuss novel approaches to mitigate the growing problem of cross-layer bugs at design time.
\end{abstract}

\section{Introduction}
The divide between hardware and software security research is starting to take its toll, as we are witnessing increasingly sophisticated attacks that are combining software and hardware bugs to exploit computing platforms at runtime~\cite{cachebleed2017yarom,memjam2018moghimi,kim2014flipping,seaborn2015exploiting,clkscrew2017tang,meltdown2018lipp,foreshadow,gras2018translation, evtyushkin2018branchscope}.
These cross-layer attacks disrupt traditional threat models which assume either hardware or software adversaries.
For instance, attacks may provoke physical effects to induce hardware faults or deliberately trigger transient microarchitectural states. Such attacks make the resulting failure modes visible to software adversaries enabling them to exploit hardware vulnerabilities remotely.
The affected targets range from low-end embedded devices to complex servers, that are hardened with architectural defenses such as data-execution prevention, supervisor-mode execution prevention, and advanced defenses such as control-flow integrity.

\noindent\textbf{Hardware vulnerabilities.}
Existing security mechanisms are completely circumvented~\cite{cachebleed2017yarom,memjam2018moghimi,seaborn2015exploiting,clkscrew2017tang,meltdown2018lipp,foreshadow,gras2018translation, evtyushkin2018branchscope} by cross-layer attacks due to the exclusive focus on mitigating attacks that exploit software vulnerabilities.
Moreover, hardware-security extensions such as Sanctum~\cite{costan2016sanctum}, Intel SGX~\cite{intel-sgx}, and ARM TrustZone~\cite{trustzone-wp} are not designed to tackle hardware vulnerabilities. Their implementation remains vulnerable to potentially undetected hardware bugs committed at design-time, and in fact, SGX and TrustZone have been targets of successful cross-layer attacks~\cite{foreshadow,clkscrew2017tang}. While Sanctum's formal model offers provable security guarantees, its trusted abstract platform model is formulated at a high level of abstraction. This approach does not ensure security at the hardware implementation level~\cite{subramanyan2017formal}.
Hardware vulnerabilities can be introduced due to: (a)~incorrect or ambiguous security specifications, (b)~incorrect design, (c)~faulty implementation of the design, or (d)~a combination thereof.
Implementation bugs occur through human error or imperfections in the high-level translation and gate-level synthesis throughout several stages of the integrated circuit (IC) design flow.
IC design is typically implemented at the register-transfer level (RTL) by hardware designers using hardware description languages (HDLs) such as Verilog and VHDL, which is \emph{synthesized} into a lower-level representation using compilers and automated tools.
Just like software programmers introducing bugs to the high-level code, hardware engineers may accidentally introduce bugs to the RTL described in this high-level HDL.
Software errors typically cause programs to crash, triggering various fallback routines to ensure the safety and security of all other programs running on the platform.
However, no such safety net exists for hardware bugs.
Therefore, even minor glitches in the implementation of, e.g., a hardware subsystem within the processor may cause the entire platform to  break down completely.\footnote{A behavior humorously hinted at in IBM System/360 machines in the form of a \emph{Halt-and-Catch-Fire} (HCF) instruction.}

\noindent\textbf{Detecting hardware security bugs.}
To detect such bugs, the semiconductor industry makes extensive use of a variety of verification and analysis techniques, such as simulation and emulation (also called dynamic verification) as well as formal verification through a wide range of tools. Examples for such industry-standard tools include Incisive~\cite{incisive}, Solidify~\cite{solidify}, Questa Simulation and Questa Formal~\cite{questaFormal}, OneSpin 360~\cite{oneSpin}, and JasperGold~\cite{jasperGold}.
Originally, the predecessors of these tools were designed for \emph{functional verification}, with security-relevant verification features being incorporated much later.
Additionally, while a rich body of knowledge exists within the software community, e.g., regarding software exploitation, and techniques to automatically detect software vulnerabilities~\cite{nielson1999principles,khedker2009data,llvm,shastry2016towards,cousot2005astree,vallee1999soot,evans2002improving}, widely applicable tools for analyzing HDLs are currently lagging behind~\cite{khattri2012hsdl,call-for-action,trippel2017tricheck}.
Consequently, the industry has moved towards a \emph{security development cycle} for hardware technologies---inspired by the security development lifecycle~\cite{howard2006security} for software and in line with the guidelines provided for hardware security development lifecycle~\cite{sandia_sdl}. 
This process incorporates a combination of many different techniques and toolsets such as RTL manual code audits, assertion-based testing, dynamic simulation, and automated security verification.
Although many functional bugs have been slipping through this process already in the past~\cite{blum1996reflections,foofbug1999intel}, it was widely believed that vulnerabilities with a severe impact on security can be prevented by the existing verification processes.
However, the recent outbreak of \emph{cross-layer bugs}~\cite{cachebleed2017yarom,memjam2018moghimi,seaborn2015exploiting,meltdown2018lipp,spectre,clkscrew2017tang,gras2018translation, evtyushkin2018branchscope, google-mediatek-cve, samsung-pagetable-cve, hp-laserjet-cve, apple-audiocodecs-cve, amazon-kindletouch-cve, dell-bios-cve, microsoft-hypervisor-cve} poses a spectrum of difficult challenges to the available security verification techniques, as these attacks exploit complex and subtle interdependencies between hardware and software.
However, existing verification techniques are limited in modeling and verifying such subtle hardware/software interactions. They currently do not scale with the size and complexity of real-world SoC designs, allowing attackers to completely disarm the basic security mechanisms of several millions of computer systems.

\noindent\textbf{Goal and Contributions.}
In this paper, we analyze the effectiveness of current hardware security verification techniques in depth.
We conducted two extensive case studies to systematically assess existing verification techniques with a strong focus on bugs committed in the RTL coding.
In joint collaboration with our industry partners, we examined public Common Vulnerabilities and Exposures (CVEs)~\cite{amd-asic-cve,amd-epyc-cve,broadcom-wifi-dos,meltdown2018lipp,spectre}, and compiled a list of over 30 RTL bugs based on real-world errata such as missing case statements, wrong if-else conditions, and flipped bus bits in parameter definitions.
Even seemingly minor RTL bugs in modern system-on-chip~(SoC) implementations can have severe consequences for security in practice due to the complex inter-dependencies between different RTL modules. 
To reproduce this effect, we implemented the list of bugs using two popular and freely available processor designs for the widely used open-source RISC-V architecture.
The open-source nature of RISC-V allows for chip designers to easily design their processor microarchitecture to implement the open-source instruction set without concerns over intellectual property, license or royalty fees.
Together with industry experts we injected the selected list of real-world bugs into the RTL code of these two processor implementations.
To evaluate how well industry-standard tools and state-of-the-art methods can detect these bugs, we then conducted our first case study for the security verification of RTL code.
Interestingly, during our in-depth analysis of the list of real-world RTL bugs, we found certain unique properties that currently pose significant and fundamental challenges for state-of-the-art verification techniques with respect to black-box abstraction, timing flow, and non-register states.
In particular, our experiments show that these fundamental challenges may cause current state-of-the-art tools to miss high-impact security bugs in the RTL code of real-world SoCs.
Often, automated verification approaches are complemented by manual inspection and code audits during the verification process in practice. Thus, we launched our second case study through the international and public hardware security competition \emph{Hack@DAC} in which 54 teams of researchers competed for three months to manually uncover the RTL coding bugs.
Our combined results from the two case studies are alarming: particular classes of hardware bugs entirely evade detection---even when complementing systematic, tool-based verification using state-of-the-art approaches with extensive manual inspection by expert teams.
Specifically, we observe that RTL bugs arising from complex and cross-modular interactions in real-world SoCs render RTL bugs extremely difficult to detect in practice. Further, it may often be feasible to exploit them from software to compromise the entire platform, and we call such bugs \emph{HardFails}.
To the best of our knowledge, we are the first to provide a systematic comparison and in-depth analysis of state-of-the-art hardware verification approaches with a focus on security-relevant bugs.
To underline the real-world security threat of HardFails, we further construct a proof-of-concept exploit based on one of our selected vulnerabilities from the Hack@DAC competition, which remotely compromises the platform in a software-only attack.
With HardFails, we systematically identify the limits of current hardware security verification approaches. Our results indicate that additional research is required to improve the state of the art in detection techniques available to the semiconductor industry.

To summarize, our main contributions are:
\begin{itemize}
    \item \textbf{Stealthy hardware bugs:} We identify \emph{HardFails} as bugs in RTL coding that are distinctly challenging to detect using existing security verification techniques and industry-leading tools.
    Besides evading detection, we highlight the gravity of these bugs by demonstrating how they remain exposed to software attackers due to the semantic gap in current hardware verification approaches. Also, we explain the fundamental limitations of current detection approaches in detail using concrete examples.
    \item \textbf{Systematic evaluation and case studies:} We compile and implement a list of RTL bugs based on real-world vulnerabilities and provide two extensive case studies on which we base our observations: (1)~An in-depth investigation of the Hack@DAC bugs using current state-of-the-art and industry-leading security formal verification tools. (2)~The Hack@DAC 2018 hardware security competition, in which 54 independent teams of researchers competed worldwide over three months to find these bugs by dynamic verification approaches such as simulation and RTL manual auditing.
    Our results are alarming and demonstrate that particular classes of bugs entirely evade detection in practice, despite extensive security verification processes that combine tool-based and manual analysis.    
    \item \textbf{Proof-of-concept exploit:} We construct an exploit based on a bug that evaded detection by all teams in the competition to demonstrate the real-world threat posed by software-based attacks in light of HardFails. We also categorize existing attacks and show that the recent outbreak of software-exploitable hardware bugs may be due to the unique properties of HardFails.
\end{itemize}

The remainder of the paper is structured as follows: in Section~\ref{sec:sdl}, we outline a typical Security Development Lifecycle adopted by the semiconductor industry.
In Section~\ref{sec:adversary}, we explain our detailed threat model.
Section~\ref{sec:bugs} summarizes and explains the main observations from our two in-depth studies by giving concrete examples.
Section~\ref{sec:limit} and Section~\ref{sec:hackdac} cover our systematic assessment of state-of-the-art formal and dynamic verification and detection approaches in detail, respectively.
In Section~\ref{sec:exploit}, we provide our proof-of-concept exploit.
Section~\ref{sec:discussion} discusses other practical issues with RTL security verification, potential mitigations, and future work.
In Section~\ref{sec:related}, we provide a comparison of the related work, to finally conclude in Section~\ref{sec:conclusion}.

\section{SoC Development Cycle and Security}\label{sec:sdl}

\begin{figure}[tb]
    \centering
    \includegraphics[width=.48\textwidth]{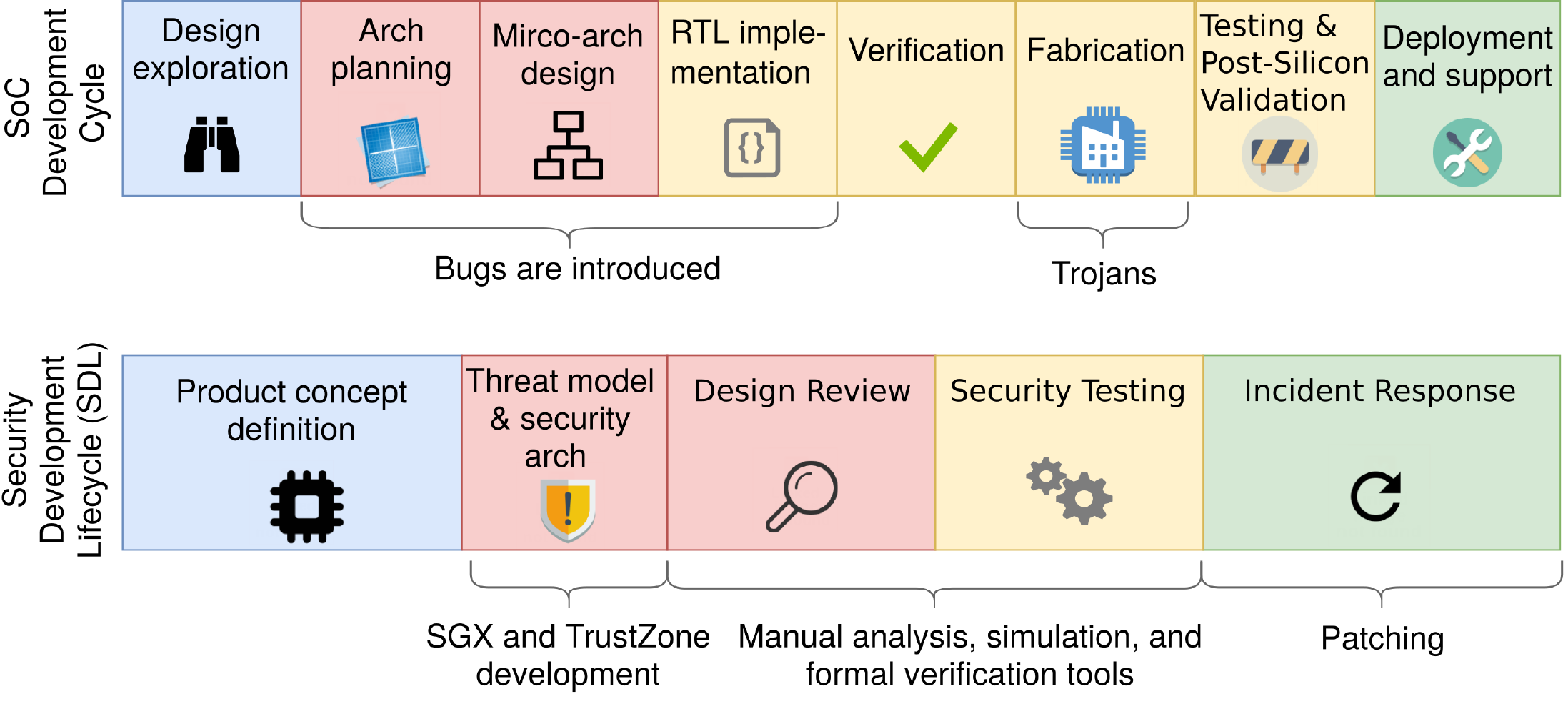}
    \caption{Typical Security Development Lifecycle (SDL) process followed by semiconductor companies.}
    \label{fig:sdl}
\end{figure}

Software companies use the Security Development Lifecycle~(SDL) process for the development of secure code \cite{howard2006security}.
This process has inspired the semiconductor companies \cite{khattri2012hsdl,cisco-hardware-sdl,lenovo-hardware-sdl} to adapt and customize the process for secure hardware design~\cite{call_for_sdl}.
This cycle overlaps with  the hardware development lifecycle~\cite{sandia_sdl}.

The top half of Figure \ref{fig:sdl} illustrates the chip development lifecyle. 
After exploration based on market research and competitive analysis, the product architecture is defined. 
This is followed by performance and power modeling on cycle-accurate simulators to optimize and finalize architectural details. 
Based on the finalized architecture, the microarchitecture is designed and implemented in RTL.
Simultaneously, pre-silicon validation efforts are undertaken to fix functional issues. 
After tape-out and fabrication, the chip is powered on and the platform bring-up step ensures that the chip is functional. 
Post-silicon validation follows, and a new stepping is spun out if necessary. 
After the production stepping passes quality requirements, the chip is shipped. 
Any issues found in the field are debugged post-mortem, and appropriate patches are released if possible, otherwise the product is recalled.

A typical SDL process followed by semiconductor vendors consists of five phases shown in the bottom half of Figure~\ref{fig:sdl}. 
After the product architectural features are finalized, a security assessment is performed based on the use case of the end product. 
This constitutes the first phase.  
In the second phase, security objectives of the chip are defined. 
A comprehensive list of assets, entry points to access those assets, adversary model, as well as all the threats are identified. 
Architectural mitigations for the threats are documented, and the security architecture is finalized. 
In the third phase, the architectural security objectives from the previous phase are translated into microarchitectural security requirements. 
Security test cases (both positive and negative) are documented. 
In the fourth phase, pre-silicon security tests are conducted using dynamic verification (i.e., simulation and emulation), as well as formal verification, which are also complemented by manual RTL reviews. 
After the chip is fabricated, post-silicon security tests are executed as part of the fifth phase using several custom or industry standard debug tools. 
The identified security bugs in both pre-silicon and post-silicon phases are rated for severity using the industry-standard scoring systems such as the Common Vulnerability Scoring System (CVSS)~\cite{cvss-specification} or in-house equivalents and are fixed promptly. 
Issues in shipping products are handled by incident response teams. 

Despite having multiple tools and validation frameworks in industry, existing solutions for detecting security vulnerabilities largely rely on human expertise to define the security test cases and run the tests. 
Even for experts, this is a tedious and highly complex task where some corner case bugs can be hard to detect using existing tools and methodologies.

\section{Adversary Model}\label{sec:adversary}
For our in-depth studies and definition of HardFails, we investigate specific microarchitectural details at the RTL level.
As all vendors keep their proprietary industry designs and implementations inaccessible, we use the popular open-source RISC-V architecture and hardware micro-architecture as a baseline~\cite{riscv}. 
RISC-V supports a wide range of possible configurations with many standard features that are also available in modern processor designs, such as privilege level separation, virtual memory, and multi-threading, as well as more advanced features such as configurable branch prediction and non-blocking data caches~\cite{rocket-core}, or out-of-order execution~\cite{celio2017boomv2}, making the platform a suitable target for our study.

RISC-V RTL is freely available and open to inspection and modification.
We note that while this is not necessarily the case for industry-leading chip designs, an adversary might be able to reverse engineer parts of the chip.
Although a highly cumbersome and difficult task in practice, this possibility cannot be excluded in principle. Hence, we allow an adversary to inspect the RTL code in our model.

In particular, we make the following assumptions to evaluate both existing verification approaches and possible attacks:

\begin{itemize}
\item \textbf{Hardware Vulnerability:} the attacker has knowledge of a vulnerability in the hardware design of the SoC (i.e., at the RTL level) and can trigger the bug from software.
\item \textbf{User Access:} the attacker has complete control over a user-space process, i.e., can issue unprivileged instructions and system calls. For RISC-V, this means the attacker can execute any instruction in the basic instruction set.
\item \textbf{Secure Software:} software vulnerabilities and resulting attacks such as code-reuse~\cite{rop-shacham,rop-wo-returns,srop,brop,rop-rootkit} and data-only attacks~\cite{dop,dataonly-attack,flowstitch,ptrand} against the software stack are orthogonal to the problem of cross-layer bugs, which leverage \emph{hardware vulnerabilities} from the software layer. In our model, all platform software could be protected by defenses such as control-flow integrity~\cite{abadi2005control} and data-flow integrity~\cite{castro2006securing}, or be formally verified.
\end{itemize}

The goal of an adversary under this model is to leverage the vulnerability on the chip to provoke unintended functionality, e.g., access to protected memory locations, code execution with elevated privileges, breaking the isolation of other processes running on the platform, or permanently denying services at the hardware level.
RTL bugs in certain modules of the chip might only be exploitable with physical access to the victim device, for instance, bugs in the implementation of debugging interfaces.
However, software-exploitable vulnerabilities can also be exploited completely remotely by software means, and hence, have a higher impact in practice.
For this reason, we focus on software-exploitable RTL vulnerabilities.
We also note that an adversary with unprivileged access is a realistic model for real-world SoCs: many platforms provide services to other devices over the local network, or even over the internet.
Consequently, the attacker can obtain some limited software access to the platform already, e.g., through a webserver or an RPC interface.

The goal of the various verification approaches in this setting is to catch all of the bugs that would be exploitable by such an adversary before the chip design enters the production phase.
In the next section, we give an overview of why current verification approaches face fundamental limitations in practice.

\section{HardFails: Stealthy Hardware Security Bugs} \label{sec:bugs}
In this section, we first explain the nature of the bugs we focus on for our in-depth investigation. We selected these bugs based on real-world hardware bugs that were previously encountered and reported in public hardware errata and CVE lists.
We then summarize our observations and explain in detail what constitutes a HardFail, by listing the explicit properties we encountered that make vulnerabilities extremely challenging to detect using state-of-the-art verification approaches. 
We then give concrete examples of such complex vulnerabilities in real-world open-source SoC RTL implementations.

\subsection{Real-World RTL Bugs}
We base our findings on investigating a solid representative spectrum of real-world RTL bugs.
Specifically, we inserted vulnerabilities inspired by recently published vulnerabilities~\cite{cachebleed2017yarom,memjam2018moghimi,seaborn2015exploiting,meltdown2018lipp,spectre,clkscrew2017tang,gras2018translation, evtyushkin2018branchscope}, security errata, and CVEs~\cite{ google-mediatek-cve, samsung-pagetable-cve, hp-laserjet-cve, apple-audiocodecs-cve, amazon-kindletouch-cve, dell-bios-cve, microsoft-hypervisor-cve} from major semiconductor manufacturing vendors and additionally new vulnerabilities specifically tailored for RISC-V. 
We investigated how these vulnerabilities can be effectively detected using formal verification techniques (Section~\ref{sec:limit}) using an industry-standard tool and in a second case study through simulation and manual RTL analysis (Section~\ref{sec:hackdac}).

Modern processors are highly complex and incorporate hundreds of different in-house and potentially also third-party Intellectual Property (IP) components. Such designs open up plenty of room for many possible pitfalls and chances for vulnerabilities being introduced in the inter-modular interactions across multiple layers of the design hierarchy. 
At the highest level, multi-core architectures typically have an intricate interconnect fabric between individual cores (implementing complicated communication bus protocols), multi-level cache controllers with shared un-core and private on-core caches, memory and interrupt controllers, as well as debug and I/O interfaces to name a few. 
For each core, these high-level components further break down to logical modules such as fetch and decode stages, an instruction scheduler, individual execution units, branch prediction, instruction and data caches, the memory subsystem, and re-order buffers, and queues. 
These are, in turn, implemented and connected using individual RTL modules. The average size of each module is several hundred code lines.
As a result, real-world SoCs can easily approach 100,000 lines of RTL code, and some open-source designs significantly outgrow this to many millions lines of code~\cite{opensparc-loc-2018}.

The majority of processors used in practice (Intel x86, AMD x86 and ARM) are based on proprietary RTL implementations that can only be licensed and partially accessed by other chip vendors. Hence,  we do not have access to their RTL implementations.  
Instead, we mimic the reported failure cases: we reproduce these bugs by injecting them deliberately into the RTL of a widely-used open-source SoC. 
Also, we investigate complex microarchitecture features of another popular open-source core and discover vulnerabilities already existing in its RTL (Section \ref{sec:ariane}). 
These RTL bugs usually manifest as:
\begin{itemize}
\item \textbf{Incorrect assignment bugs} associated with variables, registers, and parameters being assigned incorrect literal values, incorrectly connected or left floating unintended.
\item \textbf{Timing bugs} resulting from timing flow issues and incorrect behavior relevant to clock signalling such as information leakage. 
\item \textbf{Incorrect case statement bugs} in finite state machine (FSM) descriptions such as incorrect or incomplete selection criteria, or incorrect behaviour within a case.
\item \textbf{Incorrect if-else conditional bugs} associated with incorrect boolean conditions or incorrect behaviour described within either branch. 
\item \textbf{Specification bugs} associated with a mismatch between a specified property and its actual implementation or poorly specified / under-specified behavior.
\end{itemize}

What renders these seemingly minor RTL coding errors as security vulnerabilities that are also very challenging to detect during verification is how they are interconnected with the surrounding logic. This, in turn, affects the complexity of the side effects that they generate in their manifestation.
Some of these RTL bugs may be patched by adjusting parts of the software stack that uses the hardware (e.g., using firmware/microcode updates) to circumvent them and mitigate specific exploits.
However, since RTL code is usually compiled and hardwired as integrated circuitry logic, the underlying bugs will remain and cannot, in principle, be patched after production.
This is why RTL bugs pose a severe security threat in practice.

Furthermore, the limited way in which current detection approaches model hardware designs and formulate and capture security assertions raise significant challenges for current verification approaches. Such effects aggravate the impact of HardFails on real-world chip designs.

\subsection{Four HardFail Properties}
By analyzing the state-of-the-art verification tools and the results from Hack@DAC~2018, we observed four distinct properties which render RTL bugs  challenging to detect---especially when a bug exhibits multiple of these properties.
We call these the \emph{HardFail properties} of a bug:

\noindent \textbf{Cross-modular effects (HF-1).}\\
Hardware modules are often interconnected in a highly hierarchical design with multiple horizontal and vertical interdependencies. 
Thus, an RTL bug located in an individual module may trigger a vulnerability in information flow that spans multiple complex modules. 
Pinpointing the bug requires analyzing the flow across all of the relevant modules (both intra-modular and inter-modular flows). 
This is highly cumbersome and unreliable to identify by manual inspection and also quickly drives systematic formal verification approaches to their limits. 
Existing verification approaches are focused on systematic modeling and analysis of each RTL module to verify whether design specifications (expressed using security property assertions) and implementation match. 
Detecting vulnerabilities with side effects that span across multiple modules requires loading the RTL code of all the relevant modules and analyzing their intra- and inter-modular states.
Currently, these complex signal flows are quickly driving existing tools into a \emph{state explosion problem} due to the exponential relationships in the underlying modeling algorithms~\cite{clarke2012model, farahmandi2018formal}. 
While providing additional computational resources is often used as an ad-hoc solution, these resources  are usually quickly outgrown as a modeled module and flow complexity increase. 
Selective "black-box" abstraction of some of the modules, state space constraining, and bounded-model checking are some techniques that help decrease the state space explosion. However, they do not eliminate the underlying limitations. 
Additionally, these techniques introduce false negatives, and hence, they are less reliable since vulnerabilities may be missed if the black-boxing and constraining decisions are not well-reasoned by the verification engineer. 
Thus, to scale with complex SoC designs, current state-of-the-art verification approaches require interactive input and feedback from a human expert in practice.
In case the human expert's input is erroneous, the verification results are essentially void.

\noindent\textbf{Timing-flow gap (HF-2).}\\
Existing verification techniques validate security properties by checking a set of property assertions, invariants, and the absence of illegal information flows against a model of the target hardware design.
However, the currently available industry-standard approaches and tools are very limited in this respect.
This lack of expressiveness becomes especially apparent when verifying security properties related to timing flow (in terms of clock cycle latency) of the logic described in the RTL code.
In practice, this leads to vast sources of information leakage due to software-exploitable timing channels (see Section~\ref{sec:related}).
At RTL, a timing flow or channel exists from the circuit inputs to outputs when the number of clock cycles that is required for the outputs to be generated depends on the values of the inputs or the current memory/register state.
This can be exploited to leak sensitive information when the timing variation is discernible by an adversary and can be used to infer inputs or memory states.
In the RTL code, this is especially problematic for information flows and resource sharing across different privilege levels.
This timing variation should remain indistinguishable in the RTL implementation or measuring from software should be prevented.
However, current industry-standard security verification techniques focus exclusively on the functional information flow of the logic and fail to model the associated timing flow.
The complexity of timing-related security issues is further aggravated when the timing flow along a logic path spans multiple modules and involves various interdependencies.

\noindent \textbf{Cache-state gap (HF-3).}\\
Existing verification tools offer support for modeling the hardware modules and validating its security properties, e.g., in the form of assertions and invariants. Since they are mainly derived from (and still used in combination with) functional verification, they do not provide integrated support for modeling and reasoning on the properties of non-register states in the design, such as caches.
This can lead to severe security vulnerabilities arising due to state changes that are unaccounted for, e.g., cache state changes across privilege levels. 
In particular, current tools reason about the \emph{architectural state} of a processor by exclusively focusing on the state of registers.
However, this definition of the architectural state completely discards that modern processors feature a highly complex microarchitecture and diverse hierarchy of \emph{non-register} caches.
This problem is amplified as these caches have multiple levels and shared across multiple privilege levels. 
Caches represent a state that is influenced directly or indirectly by many control-path signals.
This may generate security vulnerabilities in their interactions with the processor register states, such as illegal information channel leakages across different privilege levels.
Due to these significant limitations, automatically identifying RTL bugs that trigger such vulnerabilities currently is beyond the capabilities of existing approaches.

\noindent \textbf{Hardware/firmware interactions (HF-4).}
Some RTL bugs remain indiscernible to hardware security verification techniques because they are not explicitly vulnerable unless triggered by the overlying firmware.
While many SoC access control policies are directly implemented in hardware, some of them are programmable by the firmware to allow for post-silicon flexibility.
Hence, reasoning on whether an RTL bug exists is inconclusive when considering the hardware RTL in isolation.
These vulnerabilities would only materialize when the hardware/firmware interactions are considered in combination and how the firmware programs relevant registers in the underlying hardware is modeled in the tool.
Again, we found that this type of vulnerability is largely beyond the scope of existing tools, which have little to no support for cross-verification beyond a relatively small number of cycles, which cannot account for complex real-world firmware.

\subsection{Concrete Examples for HardFails}\label{sec:ariane}
\begin{figure}[tb]
    \centering
    \includegraphics[width=.5\textwidth]{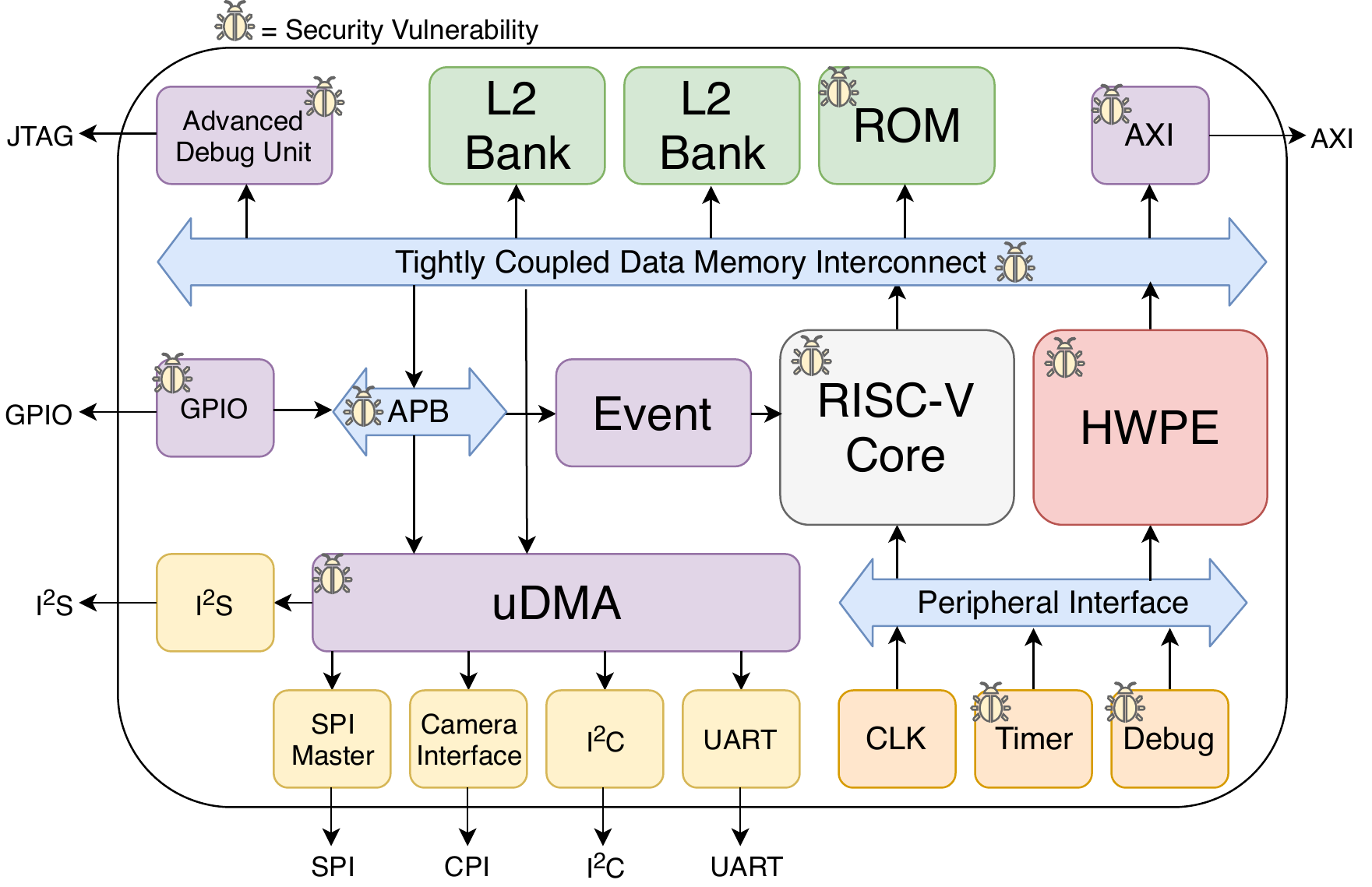}
    \caption{Hardware overview of the PULPissimo SoC.  Each bug icon indicates the presence of at least one  security vulnerability in the module.}
    \label{fig:pulpissimoAttackSurface}
\end{figure}

Next, we describe concrete examples for some of the bugs we encountered during our analysis of two different RISC-V SoCs, Ariane~\cite{ariane}
and PULPissimo~\cite{pulpissimo}. 
All these bugs are particularly challenging to detect using standard verification approaches. 
Ariane is a 6-stage in-order RISC-V CPU that implements the RISC-V draft privilege specification and can run a RISC-V Linux OS. It is implemented with a memory management unit (MMU) that consists of data and instruction transaction lookaside buffers (TLBs), a hardware page table walker, and a simple branch prediction unit to enable speculative execution.
Figure~\ref{fig:ariane_core} in Appendix~\ref{sec:ariane_figs} shows its microarchitecture. 
PULPissimo is an SoC based on a smaller and simpler RISC-V core with both instruction and data RAM (see Figure~\ref{fig:pulpissimoAttackSurface}). It provides an Advanced Extensible Interface (AXI) for accessing memory from the cores, with peripherals on an Advanced Peripheral Bus (APB) which is used to connect peripherals to the AXI through a bridge module. It provides support for autonomous I/O, external interrupt controllers and integration of hardware processing engines. It also features a debug unit and an SPI slave~\cite{pulpissimoDS}.

\noindent\textbf{TLB Page Fault Timing Side Channel (HF-1~\& \mbox{HF-2}).}\\
On analyzing the RTL of Ariane, we observed that TLB page faults due to illegal accesses occur in a different number of clock cycles than page faults that occur due to unmapped memory (we contacted the developers and they acknowledged the vulnerability).
This timing disparity in the RTL manifests in the microarchitectural behaviour of the processor. Thus, it constitutes a software-visible side channel due to the measurable clock-cycle difference of the two cases. Previous work already demonstrated how this can be exploited by user-space adversaries to probe mapped and unmapped pages.
For instance, they can be used to break randomization-based defenses~\cite{kaslr-attack,kaslr-attack-tsx,jang2016breaking,gruss2016prefetch}.
Timing-related flows can be captured by appropriate assertions that could be defined according to the security spec of the processor. However, there are significant challenges of detecting this bug in practice:
We identify at least seven RTL modules that would need to be modeled, analyzed and verified in combination, namely: ``mmu.sv'' - ``nbdcache.sv'' - 2 ``tlb.sv'' instantiations - "ptw.sv" - "load$\_$unit.sv" - ``store$\_$unit.sv''. Besides modeling their inter- and intra-modular logic flows, the timing flows would need to be modeled and tracked to formally prove absence of timing channel leakage, which is \emph{not supported} by current industry-standard tools.
Hence, the only remaining alternative is to verify this property by manually inspecting and following the clock cycle transitions. Naturally, investigating the behavior of the relevant RTL logic is highly cumbersome and error-prone. 
However, these checks must be performed, no matter how complex the design modules are.
It must be verified that timing side-channel resilience is implemented in the design (to match the security specification if required) and furthermore, that it is implemented correctly and bug-free in the RTL.
We show the RTL hierarchy of the Ariane core in Figure~\ref{fig:ariane_hierarchy} in Appendix~\ref{sec:ariane_figs} to illustrate its complexity.

\noindent\textbf{Pre-Fetched Cache State Not Rolled Back (HF-1~\&~\mbox{HF-3}).}\\
We observed another issue in Ariane with the cache state:
when a system return instruction is executed, where the privilege level of the core is not changed until this instruction is retired.
Before retirement, linear fetching (guided by branch prediction) of data and instructions following the unretired system return instruction continues at the current higher system privilege level.
Once the instruction is retired, the execution mode of the core is changed to the unprivileged level, but the entries that were prefetched into the cache (at the system privilege level) do not get flushed. While we did not construct an end-to-end attack to exploit, this and such shared cache entries are visible to user-space software, enabling timing channels between privileged and unprivileged software in principle.
Verifying the implementation of all the flush control signals and their behaviour in all different states of the processor requires examining at least eight RTL modules:  "ariane.sv" - "controller.sv" -  "frontend.sv" - "id$\_$stage.sv" - "icache.sv"  - "fetch\_fifo" - "ariane$\_$pkg.sv" - "csr$\_$regfile.sv" (see Figure~\ref{fig:ariane_hierarchy}).
This is highly complex because it requires identifying and defining all the relevant security properties to be checked across these RTL modules.
Since current industry-standard approaches do not support expressive capturing and the verification of non-register state changes, such as caches, this issue in the RTL can currently only be found by manual inspection.

\noindent\textbf{Firmware-Configured Memory Ranges (HF-4).}\\
In preparation for Hack@DAC~2018, we added peripherals to Pulpissimo and injected bugs in them to reproduce bugs from real-world hardware errata. Among the peripherals, we added is an AES encryption/decryption engine whose input key is stored and fetched from memory tightly coupled to the processor. Further detail on the bug is shown in Section \ref{sec:limit}. 
The memory address that the key is stored in is unknown, and whether it is within the protected memory range or not is inconclusive by observing the RTL alone.
In real-world scenarios, the AES key is stored in programmable fuses. Upon system start-up, the firmware has to read the key from the fuses to registers or protected memory that only the AES engine is permitted to access. The firmware would usually also program relevant registers in the underlying hardware to configure the memory address ranges and relevant access control policies. While the information flow of the AES key is strictly defined in hardware, its location is actually controlled by the firmware. Hence, reasoning on whether the information flow is allowed or not using conventional hardware verification approaches is inconclusive when considering the RTL code in isolation. The vulnerable hardware/firmware interactions cannot be identified unless both the hardware and the firmware are co-verified. Unfortunately, current industry-standard tools do not support such an analysis.

\noindent\textbf{Memory Address Range Overlap (HF-1 \& HF-4).}\\
Pulpissimo provides I/O support to its peripherals by mapping them to different memory address ranges. If an address range overlap bug is accidentally committed at design-time or by the firmware, this can break access control policies and have critical security consequences, e.g., privilege escalation. We injected an RTL bug where there is address range overlap between the SPI Master Peripheral and the SoC Control Peripheral which allowed the untrusted SPI Master to access the SoC Control memory address range over the AMBA APB bus. 
Verifying issues at the SoC interconnect in complex ARM bus protocols is highly challenging since too many modules needed to support the interconnect have to be modeled to properly test the bug. This greatly increases the scope and the complexity of the bug far beyond just a few modules.
Such an effect causes an explosion of the state space, since all the possible states have to be modeled accurately to remain sound.
While proof kits for accelerated certification of advanced SoC interconnect protocols were introduced to mitigate this effect for a small number of bus protocols specifically (here, AMBA3 and AMBA4), this requires an add-on to the default software and many protocols are not supported~\cite{intelligentProofKits}.

\section{In-depth Study of Detection of HardFails}\label{sec:limit}
In practice, hardware-security verification engineers use a combination of techniques such as formal verification, simulation, emulation, and manual inspection.  We focus in our first case study (and in this section) on evaluating the effectiveness of industry-standard formal verification techniques used for detecting hardware security bugs. The next section describes the effectiveness of simulation and manual inspection techniques used by the teams in our competition.
We emphasize that in a real-world security testing (see Section~\ref{sec:sdl}), engineers will not have prior knowledge of the specific vulnerabilities they are trying to find.
Our goal, however, is to investigate how an industry-standard tool can detect RTL bugs that we deliberately inject in an open-source SoC and have prior knowledge of (see Table~\ref{tab:pulpissimo-JG-Results}).
We then analyze our results in this controlled setting to identify why and how current tools fail to detect these bugs.

\begin{table*}[tp]
\normalsize
\setstretch{1}
\centering
\resizebox{1\textwidth}{!}{
	\begin{tabular}{@{}c l c c c c c c@{}}
		\toprule
		\# & \textbf{Bug} & \textbf{Type} & \textbf{SPV} & \textbf{FPV} & \begin{tabular}{@{}l@{}}\textbf{Hack@DAC}\end{tabular}
		& \begin{tabular}{@{}l@{}}\textbf{Modules}\end{tabular}
		& \textbf{Lines of code}
                \\
		\midrule
                
                1 & \begin{tabular}{@{}l@{}}
                Address range overlap between peripherals SPI Master and SoC.\end{tabular}
                & Inserted & \cmark & \cmark & \cmark
                & 91 & 6685
                \\ \\
                
                2 & \begin{tabular}{@{}l@{}}
                Addresses for L2 memory is out of the specified range.\end{tabular}
                & Native & \cmark &\cmark & \cmark
                & 43 & 6746
                \\ \\
                
                3 & \begin{tabular}{@{}l@{}}
                Processor runs code on incorrect privilege level for the CSR.\end{tabular}
                & Native & \xmark & \cmark & \cmark
                & 2 & 1186
                \\ \\
                
                4 & \begin{tabular}{@{}l@{}} 
                Register that controls GPIO lock can be written to with software.\end{tabular}
                & Inserted & \cmark & \cmark & \xmark
                & 2 & 408
                \\ \\
                
                5 & \begin{tabular}{@{}l@{}} Reset clears the GPIO lock control register.\end{tabular}
                & Inserted & \cmark & \cmark & \xmark
                & 2 & 408
                \\ \\
                
                6 & \begin{tabular}{@{}l@{}} Incorrect address range for APB allows memory aliasing.\end{tabular}
                & Inserted & \cmark & \cmark & \xmark
                & 1 & 110
                \\ \\
                
                7 & \begin{tabular}{@{}l@{}} AXI address decoder ignores errors.\end{tabular}
                & Inserted & \xmark & \cmark & \xmark
                & 1 & 227
                \\ \\
                
                8 & \begin{tabular}{@{}l@{}} 
                Address range overlap between GPIO, SPI, and SoC control peripherals.\end{tabular}
                & Inserted & \cmark & \cmark & \cmark
                & 68 & 14635
                \\ \\
                
                9 & \begin{tabular}{@{}l@{}} 
                Incorrect password checking logic in debug unit.\end{tabular}
                & Inserted & \xmark & \cmark & \xmark
                & 4 & 436
                \\ \\
                
                10 & \begin{tabular}{@{}l@{}}  Advanced debug unit only  checks 31 of the 32 bits of the password.\end{tabular}
                & Inserted & \xmark & \cmark & \xmark
                & 4 & 436
                \\ \\
                
                11 & \begin{tabular}{@{}l@{}} Able to access debug register  when in halt mode.\end{tabular}
                & Native & \xmark & \cmark & \cmark 
                & 2 & 887 
                \\ \\
                
                12 & \begin{tabular}{@{}l@{}} Password check for the debug unit does not reset after successful check.\end{tabular}
                & Inserted & \xmark & \cmark & \cmark
                & 4 & 436
                \\ \\
                
                13 & \begin{tabular}{@{}l@{}} Faulty decoder state machine logic in RISC-V core results in a hang.\end{tabular}
                & Native & \xmark & \cmark & \cmark
                & 2 & 1119
                \\ \\
                
                14 & \begin{tabular}{@{}l@{}}
                Incomplete case statement in ALU can cause unpredictable behavior.\end{tabular}
                & Native & \xmark & \cmark & \cmark
                & 2 & 1152
                \\ \\
                
                15 & \begin{tabular}{@{}l@{}}
                Faulty timing logic in the RTC results in inaccurate calculation of time.\end{tabular}
                & Native & \xmark & \cmark & \xmark
                & 1 & 191
                \\ \\
                
                16 & \begin{tabular}{@{}l@{}} Reset for the advanced debug unit not operational.\end{tabular}
                & Inserted & \xmark & \xmark & \cmark
                & 4 &  436
                \\ \\
                
                17 & \begin{tabular}{@{}l@{}} Memory-mapped register file allows code injection.\end{tabular}
                & Native & \xmark & \xmark & \cmark
                & 1 & 134
                \\ \\
                
                18 & \begin{tabular}{@{}l@{}} Non-functioning cryptography module causes DOS.\end{tabular}
                & Inserted & \xmark & \xmark & \xmark
                & 24 & 2651 
                \\ \\
                
                19 & \begin{tabular}{@{}l@{}} Insecure hash function in the cryptography module.\end{tabular}
                & Inserted & \xmark & \xmark & \xmark
                & 24 & 2651 
                \\ \\
                
                20 & \begin{tabular}{@{}l@{}} Cryptographic key for AES stored in  unprotected memory.\end{tabular}
                & Inserted & \xmark & \xmark & \xmark
                & 57 & 8955 
                \\ \\
                
                21 & \begin{tabular}{@{}l@{}} Temperature sensor is muxed with the cryptography modules.\end{tabular}
                & Inserted & \xmark & \xmark & \cmark
                & 1 & 65 
                \\ \\
                
                22 & \begin{tabular}{@{}l@{}} ROM size is too small preventing execution of security code.\end{tabular}
                & Inserted & \xmark & \xmark & \cmark
                & 1 & 751 
                \\ \\
         
                23 & \begin{tabular}{@{}l@{}} Disabled zero RISC-V core.\end{tabular}
                & Inserted & \xmark & \xmark & \xmark
                & 1 & 282 
                \\ \\
                
                24 & \begin{tabular}{@{}l@{}} GPIO enable always high.\end{tabular}
                & Inserted & \xmark & \xmark & \xmark
                & 1 & 392 
                \\ \\
                
                25 & \begin{tabular}{@{}l@{}} Secure mode not required to write to RISC-V core control registers.\end{tabular}
                & Inserted & \xmark & \xmark & \cmark
                & 1 & 745 
                \\ \\
                
                26 & \begin{tabular}{@{}l@{}} Advanced debug unit password is hard-coded and set on reset.\end{tabular}
                & Inserted & \xmark & \xmark & \cmark
                & 1 & 406 
                \\ \\
                
                27 & \begin{tabular}{@{}l@{}} Secure mode is not required to write to interrupt registers.\end{tabular}
                & Inserted & \xmark & \xmark & \cmark
                & 1 & 303 
                \\ \\
                
                28 & \begin{tabular}{@{}l@{}} JTAG interface is not password protected.\end{tabular}
                & Native & \xmark & \xmark & \cmark
                & 1 & 441 
                \\ \\
            
                29 & \begin{tabular}{@{}l@{}} Output of MAC is not erased on reset.\end{tabular}
                & Inserted & \xmark & \xmark & \cmark
                & 1 & 65 
                \\ \\

                30 & \begin{tabular}{@{}l@{}} Supervisor mode signal of a core is floating preventing the use of SMAP.\end{tabular}
                & Native & \xmark & \xmark & \cmark
                & 1 & 282 
                \\ \\
                
                31 & \begin{tabular}{@{}l@{}} GPIO is able to read/write to instruction and data cache.\end{tabular}
                & Native & \xmark & \xmark & \cmark
                & 1 & 151 
                \\

		\bottomrule
	\end{tabular}
}
	\caption{Detection results based on formal verification (\textbf{SPV} and \textbf{FPV}), and manual inspection combined with simulation~(\textbf{Hack@DAC}). Check and cross marks indicate detected and undetected PULPissimo SoC bugs respectively.}
	\label{tab:pulpissimo-JG-Results} 
\end{table*}

\subsection{Detection Methodology}
We examined each of the injected bugs and its nature in order to determine which formal technique would be best suited to detect it.
Our results in this study are based on two formal techniques: Formal Property Verification (FPV) and Security Path Verification (SPV)~\cite{JG-SPV}. They represent the state of art in hardware security verification routinely used throughout the semiconductor industry~\cite{sandia2014veriftools}.
\textbf{FPV} checks whether a set of security properties, usually specified as SystemVerilog Assertions~(SVA), hold true for the given RTL. 
To describe our assertions correctly, we examined the location of each bug in the RTL and how it is manifested in the behavior of the surrounding logic and input/output relationships. Once we specified the security properties using \emph{assert}, \emph{assume} and \emph{cover} statements, we determined which RTL modules we need to model to prove these assertions. If a security property is violated, the tool generates a counterexample; this example is examined to ensure whether the intended security property is indeed violated or a false alarm.

{\bf SPV} is used to detect bugs which specifically involve unauthorized information flow. Such properties cannot be directly captured using SVA/PSL assertions.
SPV uses path sensitization techniques to exhaustively and formally check if unauthorized data propagates (through a functional path) from a source to a destination signal. 
To specify the SPV properties, we identified source signals where the sensitive information was located and destination signals where it should \emph{not} propagate. 
We then identified the bounding preconditions to constrain the paths that the tool searches. 
Similar to FPV, we identified the modules that are required to capture the  information flow. This  includes source and destination modules, intermediate modules, and modules that generate control signals which interfere with the information flow. 
While it is simpler to include all design RTL modules, this often leads to a memory-usage explosion and is not practical for more complex SoC designs. It is necessary to select which modules are relevant for the properties being tested and which can be safely black-boxed which is time-consuming, error-prone and requires expertise. On the other hand, black-boxing introduces the possibility of false negatives and unreliable results. The absence of a counterexample to an assertion or flow property is inconclusive as to whether the assertion is indeed not violated or if the vulnerability is missed due to incorrect abstraction.

\begin{figure}[tpb]
    \centering
    \includegraphics[width=.5\textwidth]{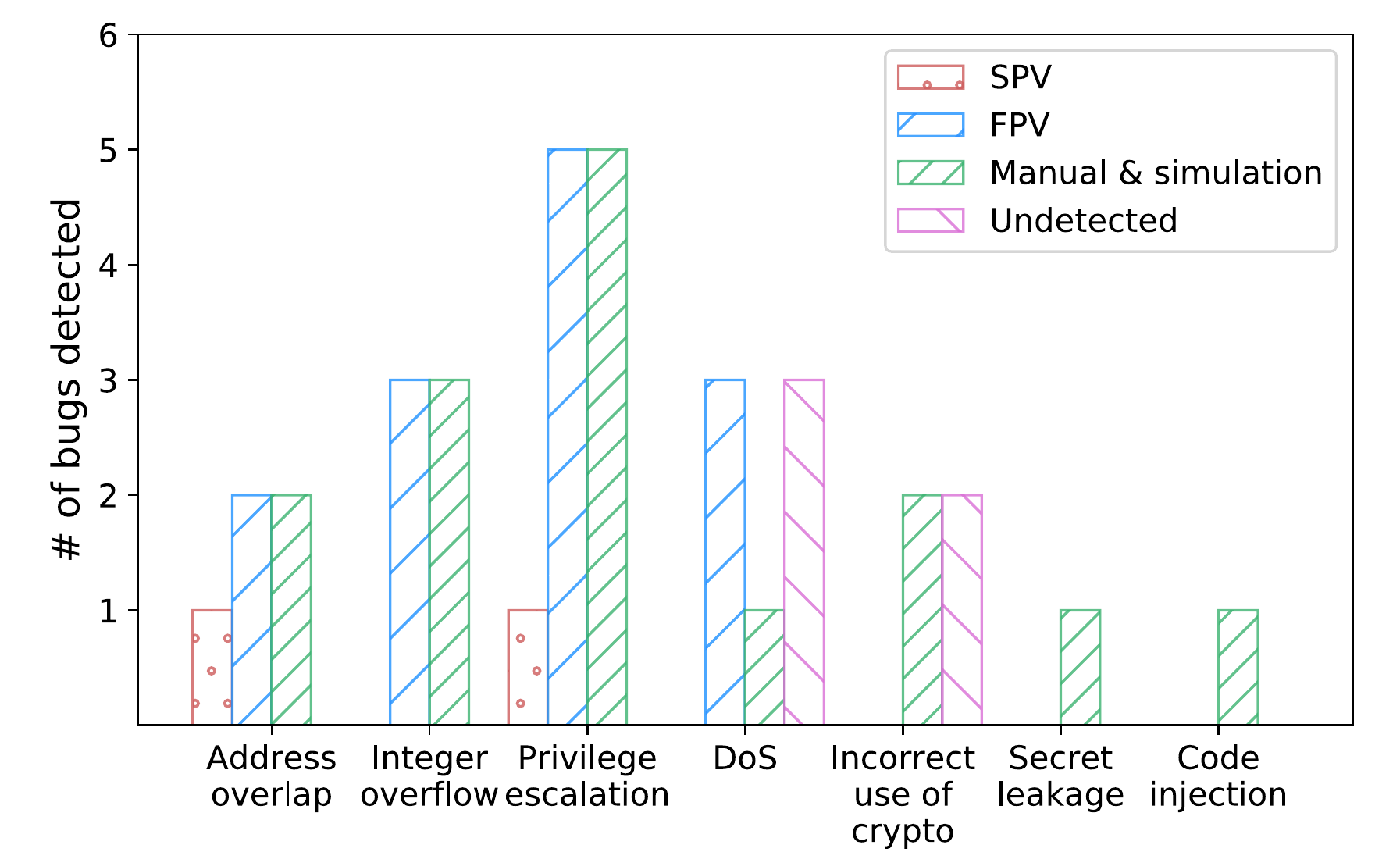}
    \caption{Verification results grouped by bug class.}
    \label{fig:jg_results}
\end{figure}

\subsection{Detection Results}
Out of the 31 bugs we investigated, shown in Table~\ref{tab:pulpissimo-JG-Results}, using the formal verification techniques described above, only 15 or 48\%, were detected.
While we tried to detect all 31 bugs formally, we were only able to formulate security properties for only 17 bugs.
This indicates that the main challenge with using formal verification tools is identifying and expressing security properties that the tools are capable of capturing and checking. 
Bugs due to ambiguous specifications of interconnect logic, for instance, are examples of bugs that are difficult to create security properties for.
Our results, shown in the SPV and FPV bars of Figure~\ref{fig:jg_results}, indicate that integer overflow and address overlap bugs had the best detection rates, 80\% and 100\%, respectively.
These classes of bugs typically involved a variable being assigned a value that is outside of the range documented in the specification, which is trivial to detect using an assertion.
For privilege escalation and denial-of-service (DoS) bugs, the detection rate is only 50\%, while secret leakage and incorrect use of crypto bugs went undetected.
The implications of these findings are especially grave for real-world more complex SoC designs where these bug classes are highly relevant from a security standpoint.

\subsection{Limitations and Challenges}
In real-world security testing, detecting all the bugs without prior knowledge of them (with knowledge only of the adversary model and security specifications) would be significantly more challenging.  
However, we assume prior knowledge of the bugs in this investigation, since we aim to shed light on the limitations of the tools.
We discuss in detail below some of the bugs that were more challenging for us to detect.

\indent{\textbf{Bug \#20: Incorrect use of crypto.}}
As mentioned in Section~\ref{sec:ariane}, the AES unit added to the PULPissimo stores its cryptographic key in the unprotected memory. This is not detectable by current formal verification tools because the address where the key is stored at is unknown to the verification engineer since it is determined by the firmware. This prevents the verification engineer from writing an assertion capable of detecting the bug using FPV or SPV. 

\indent{\textbf{Bugs \#1 and \#2: Memory address range overlap and data overflow.}}
The address range of the L2 memory implemented in the RTL does not match its specification in the SoC documentation. According to the specifications, the range should be $0x1C00\_0000$ to $0x1C08\_0000$.
However, in the RTL the address range is $0x1C01\_0000$ to $0x1C08\_2000$, as shown in Listing 1. This range does not overlap with the memory maps of other parts of the SoC but does decrease the total space mapped to the L2 memory.
To detect this bug, we wrote an assertion, shown in Listing 2, to check if the address used by the Tightly-Coupled Data Memory (TCDM) is within the range specified by the documentation.
Using this assertion, with FPV we should be able to detect this bug.
However, we have had difficulty loading the environment for this test due to the large number of modules needed to test this bug:
The bug resides at the interconnect level of the SoC. Therefore, all the modules needed to support the interconnect are also needed to properly test the bug.
This greatly increases the scope and thus causing an explosion of state space since all possible states have to be modeled accurately to remain sound.

\indent{\textbf{Bugs \#25 and \#27: Privilege Escalation.}}
We also inserted several trivial bugs. For instance, we replaced the PULP\_SECURE variable, which controls access privileges to the registers, with the PULP\_SEC variable.
While only a minor change, the results are critical for security: PULP\_SEC is a hardwired constant (i.e., it is always true). Hence, secure mode is not required to write to the interrupt control registers of the core any longer. While conceptually very simple, this is a realistic bug that could be exploited easily by attackers, e.g., by installing a malicious interrupt handler. Neither is this type of bug detectable by current formal verification tools nor was it found by any team through manual inspection and dynamic testing.
Interestingly, current tool-based approaches seem to miss such bugs because they cannot handle multiple declarations in the RTL code.
We describe some of the  bugs we inserted in greater detail in Appendix~\ref{sec:apdx_pulpissimo}.

\begin{lstlisting}[language=Verilog, caption=\textbf{Data overflow RTL:} The L2 memory address range is defined incorrectly.]
localparam logic [ADDR_WIDTH-1:0] TCDM_START_ADDR = {32'h1C01_0000}; // Start of L2 interleaved
localparam logic [ADDR_WIDTH-1:0] TCDM_END_ADDR   = {32'h1C08_2000}; // END of L2 interleaved
\end{lstlisting}
\begin{lstlisting}[language=Verilog, caption=\textbf{Data overflow assertion:} The RTL shown in listing 1 is checked to see if the address signal of the TCDM is within the range specified by the documentation.]
always @ (posedge clk) begin
	for(i=0; i<5; i++) begin  
		if (TCDM_data_gnt_DEM_TO_XBAR[i] == 1'b1) begin   
			a_addr_range_check: assert ((TCDM_data_add_DEM_TO_XBAR[i] >= 32'h1C00_0000) && (TCDM_data_add_DEM_TO_XBAR[i] <= 32'h1C08_0000)); 
		end
	end
end
\end{lstlisting}

\section{Crowdsourcing Detection}\label{sec:hackdac}
We present next the results of our second case study.
54 teams of researchers participated in Hack@DAC 2018, a recently conducted capture-the-flag competition to identify hardware bugs that were injected deliberately in real-world open-source SoC designs. 
This is the equivalent of bug bounty programs that semiconductor companies offer \cite{intel_bug_bounty,qualcomm_bug_bounty,samsung_bug_bounty,apple_bug_bounty}.
The teams were free to use any testing techniques. However, they all eventually relied on simulation and manual inspection methods, because they are easier, more accessible, and require less expertise than formal verification, especially when working under time constraints.
We injected the bugs in a joint collaboration with our industry partner inspired by their hardware security expertise.
Specifically, some bugs mimic real-world errata and publicly reported vulnerabilities from CVE lists to reproduce realistic bugs that were previously encountered.
The goal is to investigate how well these bugs can be detected through dynamic verification and manual RTL audit without prior knowledge of the bugs.
The competition consisted of two phases: a preliminary Phase 1 and final Phase 2 which featured the RISC-V Pulpino and Pulpissimo SoCs, respectively. 

\subsection{Hack@DAC 2018 Goals}
To prepare the SoCs for the competition, we first implemented additional security features into them, then defined the security objectives and adversary model and accordingly inserted the bugs into them. Specifying the security objectives and the adversary model will enable the teams to identify what would be defined as a security bug.

\noindent \textbf{Security Features:} We added password-based locks on the JTAG modules of both SoCs and access control on certain peripherals. For the Phase-2 SoC, we also added a cryptography unit implementing multiple cryptographic algorithms. We injected bugs into these features and native features to generate security threats as a result.

\noindent \textbf{Security Objectives:}
We provided the three main security objectives of the target SoCs to the teams. Firstly, an unprivileged code should not  escalate beyond its privilege level. Secondly, the JTAG module should be protected against an adversary with physical access. Finally, the SoCs should thwart software adversaries launching denial of service attacks.

\subsection{Hack@DAC 2018 Overview}
For the first phase of the competition, we chose the Pulpino SoC since it was a real-world yet not overly complex SoC design for the teams to work with. It features a RISC-V simple core with both instruction and data RAM, an AXI interconnect for accessing memory, with peripherals on an APB able to access the AXI through a bridge module. It also features a boot ROM to store boot code, a debug unit and a serial peripheral interface (SPI) slave~\cite{pulpinoDS}. We inserted security bugs in multiples modules of the SoC, including the AXI, the APB, debug unit, GPIO, and the bridge.

For the second phase, we chose the Pulpissimo SoC~\cite{pulpissimoDS}, shown in Figure~\ref{fig:pulpissimoAttackSurface}, since it supported integrating hardware processing engines, a new input/output subsystem (UDMA), and more peripherals. This allowed us to extend the SoC with additional security features, making room for inserting more bugs in them. Bugs were also inserted into the native features of the SoC, while native security bugs were also discovered afterwards. We describe some of these bugs below (more details in Appendix~\ref{sec:apdx_pulpissimo}).

\begin{itemize}
 \item \textbf{UDMA address range overlap:} We modified the memory range of the UDMA implementation so that it overlapped with the master port to the SPI. This bug allows an adversary with access to the UMDA memory to escalate its privileges and modify the memory of the SPI. 
\item \textbf{GPIO address range overlap:} The address range of the GPIO memory was erroneously declared. An adversary with GPIO access can escalate its privilege and access the SPI Master and SoC Control. 
\item \textbf{Error in GPIO Status:} GPIO enable was rigged to display a fixed erroneous status of '1', which did not give the user a correct display of the actual GPIO status.
\item \textbf{Untrusted Boot ROM:} A native bug in the SoC would allow unprivileged compromise of the boot ROM and potentially the execution of untrusted boot code at a privileged level, allowing exfiltration of sensitive information.
\item \textbf{Erroneous AXI Finite-State Machine:} 
We injected a bug in the AXI address decoder module such that if an error signal is generated on the memory bus while the underlining logic is still handling an outstanding transaction, the next signal to be handled will instead be considered operational by the module unconditionally. This bug can be exploited to intentionally cause computational faults in the execution of security critical code (we outline how to exploit this vulnerability---which was not detected by all teams---in Section~\ref{sec:exploit}).
\end{itemize}

\subsection{Hack@DAC 2018 Results}
We were able to draw various insights from the bug reports submitted by all the competitors and we indicate the results in the \emph{Manual \& Simulation} column in Table~\ref{tab:pulpissimo-JG-Results}.\\
\noindent {\bf Analyzing the Bug Reports:}
The bug reports submitted by teams provided insight into what types of bugs are harder to detect using existing approaches and which modules are harder to analyze.
Together with our industry experts, we scanned the submissions and rated the bug submissions on the accuracy and detail provided by the teams, e.g., bug validity, the methodology used, and the security impact.

\noindent {\bf Detected Bugs:}
There were two highly detected bugs in Pulpissimo. The first was a bug where the debug IPs were used when not intended, due to our added security parameters. The second bug was where we declared a local parameter PULP\_SEC, which was always set to '1', instead of the intended PULP\_SECURE. The former was detected because debugging interfaces represent security-critical regions of the chip. The latter was detected because it indicated intuitively that exploiting this parameter would lead to privilege escalation attacks. Hence, the teams prioritized inspecting these modules during the competition.

\noindent {\bf Undetected Bugs:} Many of the inserted bugs were not detected.
One was in the advanced debug unit, where the password bit index register has an overflow (bug \#9). This is an example of a security flaw that would be hard to detect by methods other than verification. Moreover, the presence of a lot of other bugs within the advanced debug unit password checker further masked this bug.
Another bug was that of the cryptographic unit key storage in unprotected memory (bug \#20). Current formal verification approaches cannot detect this bug.
By manual inspection, the teams could not also detect this bug as they focused exclusively on the RTL code in isolation and did not consider HW/FW interactions.

\noindent{\bf HardFails and Limitations of Manual Analysis:}
While manual analysis can detect the widest array of bugs, our analysis of the competition results reveals its limitations.
Manual analysis is qualitative and is difficult to scale to cross-layer and more complex bugs.
In Table~\ref{tab:pulpissimo-JG-Results}, there are 16 cross-module bugs (spanning more than one module), and only 9 of which were identified using manual inspection in the competition. We see that three of these bugs (18, 19, and 20) were also undetected by formal verification methods, which is 10\% of the bugs we investigated in our case studies. In the following section, we show how a HardFail can be easily exploited by software means to bypass the security of an SoC.

\section{Exploiting Hardware Bugs From Software} \label{sec:exploit}
We show how selected hardware bugs from Hack@DAC~2018 can be used to craft a real-world exploit. Such exploits allow unprivileged attackers to undermine the entire system by escalating privileges in an entirely remote setting.
\begin{figure}[tb]
    \centering
    \includegraphics[width=.45\textwidth]{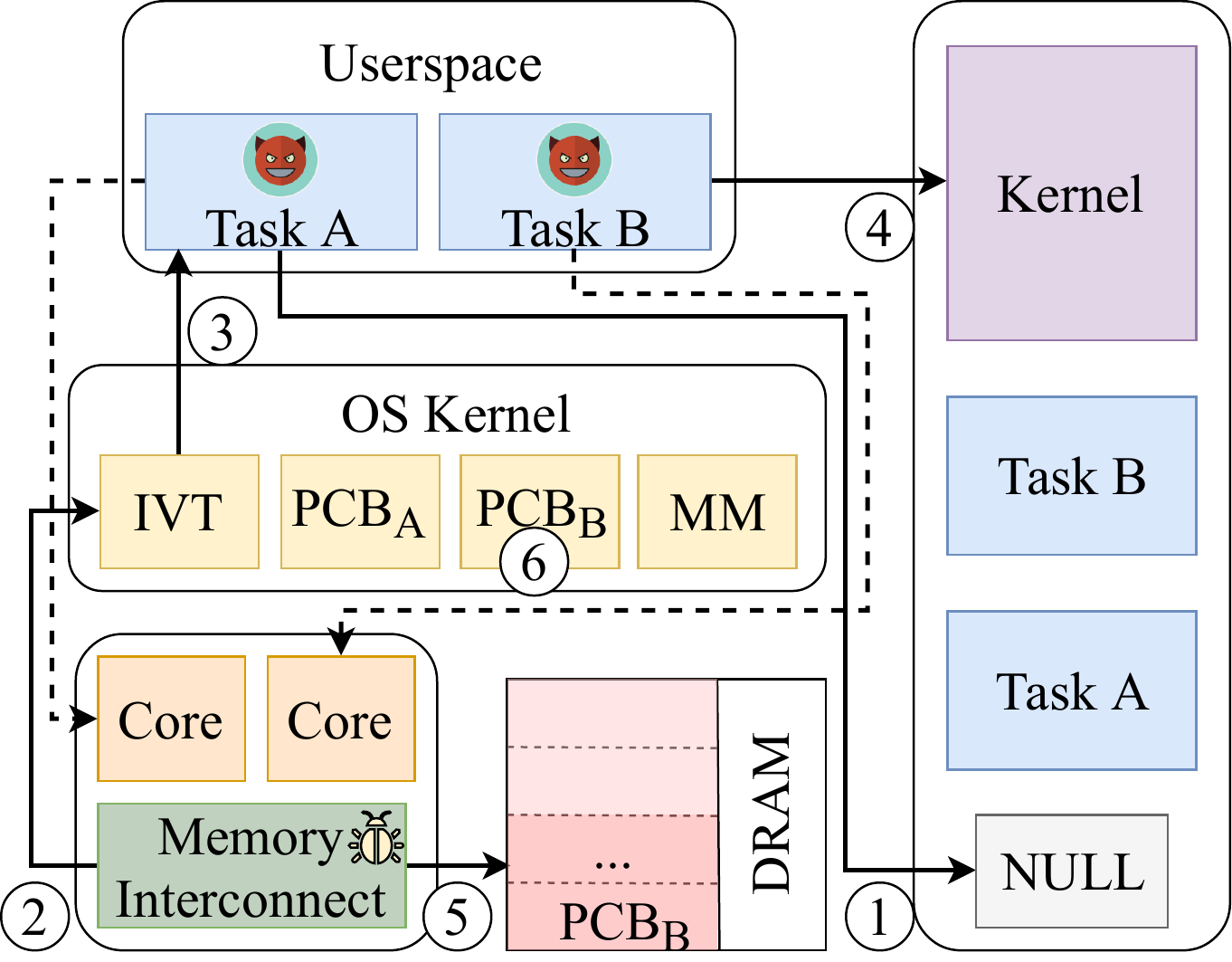}
    \caption{Our attack exploits a bug in the implementation of the memory bus of the PULPissimo SoC: by \circledw{1} \emph{spamming} the bus with invalid transactions an adversary can make \circledw{4} malicious write requests be set to \texttt{operational}.}
    \label{fig:attack}
\end{figure}
The attack is depicted in \autoref{fig:attack} in which we assume the memory bus decoder unit (unit of the memory interconnect) to have a bug, which causes errors to be ignored under certain conditions (see bug number $\#7$ in \autoref{tab:pulpissimo-JG-Results}).
This RTL vulnerability manifests in the hardware behaving in the following way. When an error signal is generated on the memory bus while the underlining logic is still handling an outstanding transaction, the next signal to be handled will instead be considered operational by the module unconditionally.
This represents a severe vulnerability, as it allows erroneous memory accesses to slip through hardware checks at runtime.
Despite this fact, we only managed to detect this vulnerability after significant efforts using FPV based on our prior knowledge of the exact location of the vulnerability.
Additionally, the tool-based (but interactive) verification procedure represented a significant costly time investment.
Since vulnerabilities are usually not known a priori in practice, this would even be more challenging.
Therefore, it is easily conceivable and realistic to assume that such a vulnerability could slip through verification and evade detection in larger real-world SoCs.

Armed with the knowledge about this vulnerability in a real-world processor, an adversary could now force memory access errors to slip through the checks as we describe in the following. In the first step \circledw{1}, the attacker generates a user program (Task A) that registers a dummy signal handler for the segmentation fault (SIGSEGV) access violation. This first program then executes a loop with \circledw{2} a faulting memory access to an invalid memory address (e.g., $LW x5, 0x0$). This will generate an error in the memory subsystem of the processor and issue an invalid memory access interrupt (i.e., \texttt{0x0000008C}) to the processor. The processor raises this interrupt to the running software (in this case the OS), using the pre-configured interrupt handler routines in software. The interrupt handler in the OS will then forward this as a signal to the faulting task \circledw{3}, which keeps looping and continuously generating invalid accesses.
Meanwhile, the attacker launches a separate Task B, which will then issue single memory access \circledw{4} to a privileged memory location (e.g., $LW x6, 0xf77c3000$).

In this situation, multiple outstanding memory transactions will be generated on the memory bus; all but one of which the address decoder will signal an error.
An invalid memory access will always proceed the single access of the second task. Due to the bug in the memory bus address decoder, \circledw{5} the malicious memory access will become \texttt{operational} instead of triggering an error.
As a result, the attacker can issue read and write instructions to arbitrary privileged (and unprivileged) memory by forcing the malicious, illegal access with preceding faulty access.
Using this technique the attacker can eventually leverage this read-write primitive, e.g., \circledw{6} to escalate privileges by writing the process control block ($PCB_{B}$) for his task to elevate the corresponding process to root.
This bug leaves the attacker with access to a root process, gaining control over the entire platform and potentially compromising all the processes running on the system.

\section{Discussion and Future Work}\label{sec:discussion}
We discuss next why microcode patching is not sufficient for RTL bugs. 
While this emphasizes the necessity of advancing pre-silicon security verification tools, we also shed light on the additional challenges associated with the hardware security verification process altogether and how it can be further improved.

\subsection{Microcode Patching}
While existing industry SoCs support hot-fixes by \emph{microcode patching}, this approach is inherently limited to a handful of changes to the instruction set architecture, e.g., modifying the interface of individual complex instructions and adding or removing instructions.
Thus, such patches at this higher abstraction level in the firmware only act as a "symptomatic" fix that circumvent the RTL bug. But they are not able to solve the fundamental problem in the RTL implementation, which is usually realized as hardwired circuits. Therefore, microcode patching is a fallback for RTL bugs discovered after production when it is too late by then to patch the RTL. They also usually come at the cost of a significant performance impact that may be avoided altogether if the underlying problem is discovered and fixed pre-silicon.

\subsection{Additional Challenges in Practice}
Our findings listed in Section \ref{sec:bugs} are based on our investigation on the efficacy of using industry-standard tools for detecting hardware security bugs.
In practice, there are additional challenges to that affect both the difficulty of detecting HardFails and their impact.

\noindent \textbf{IP Reuse.}
Some HardFails arise when the RTL code base for one product is re-purposed for a different product that has a very different set of security requirements and usage scenario. This is the very nature of hardware design and IP reuse which introduces challenges in replicating the security verification process. Parameters may be declared multiple times within this new product; they get misinterpreted by industry-standard tools, causing bugs to go undetected.

\noindent \textbf{Functional vs. Security Specifications.}
When designing hardware, the system implementation naturally deviates from its product specification, especially when system complexity increases and specification ambiguity arises.
Ideally product specification and its implementation must fully match by the time the product is ready for deployment.
This is typically accomplished through pre-silicon and post-silicon verification efforts.  
This deviation in the implementation is a result of several functional and security bugs.
A functional bug is a violation of the functional specification, that generates an incorrect result. 
These bugs are typically detected when validating the implementation against functional objectives as detailed out in a functional test plan.
Security bugs or security vulnerabilities involve unconsidered scenarios and corner cases within the specification that make the product vulnerable to attacks.  
Often, several functional bugs can also be chained to create a security bug. 
These are typically detected when validating the system implementation against product security objectives laid out in a security test plan, which is derived from the threat model under consideration. 
It is important, in practice, to clearly distinguish between functional and security specifications since these are often the references for different verification teams.

\noindent \textbf{Specification Ambiguity.}
One of the biggest challenges in practice is anticipating and identifying all the security properties that are required in a real-world scenario. We analyzed the efficacy of industry-standard tools in a controlled setting---where we purposefully inject selected bugs and have prior knowledge of them.
However, in practice this is not the case: hardware validation teams do not have prior knowledge of the bugs that they need to detect. Security specifications are often incomplete and ambiguous, often only outlining the required security properties under an assumed adversary model.
These would not hold anymore once the adversary model is changed. 
Furthermore, specs usually do not specify bugs and information flows that are \emph{not allowed} to exist and there is no automated or systematic approach to reason whether one is in fact proving the intended properties.
This can be alleviated to some extent by introducing machine-readable specifications~\cite{armv8-xml-release}. 
However, specification ambiguity can cause information leakage from a combination of incomplete or incorrect design decisions and implementation errors.

\subsection{Improving Detection}
The state of the art in security verification currently relies on repurposed tools from functional verification. 
These are a small number of detection techniques that fail to adequately model many types of vulnerabilities.
Although manual code inspection is generally useful and can potentially cover a wide array of bugs, its quality and efficacy depend exclusively on the engineer conducting the RTL audit. This is inefficient and unreliable in light of rapidly evolving and constantly growing chip designs.
Furthermore, exhaustive testing of specifications through simulation requires exponential amounts of resources in the size of the input (i.e., RTL code) while coverage must be intelligently maximized. Hence, current approaches face severe scalability challenges, as diagnosing software-exploitable bugs that reside deep in the design pipeline can require simulation of trillions of cycles~\cite{boom-loc-2016} in practice.

During our investigation of the RTL vulnerabilities, we noticed that it would often be beneficial to first identify areas with high risks due to software exposure, such as password checkers, crypto cores, and control registers, and then test them with a higher priority (this was also noted by some of the teams).

Scalability due to complex interdependencies among modules is one of the challenges in detection.  Vulnerabilities associated with non-register states (such as caches) or clock-cycle dependencies (i.e., timing flows) are also another problem.
Initial research is underway~\cite{tortuga-unison} to analyze a limited amount of low-level firmware running on top of a simulated RTL design for information and timing flow violations. However, these approaches are still at its infancy, not yet widely available and of questionable scalability for real-world SoC designs.
Finally, current verification approaches focus on register-state based analysis, e.g., to monitor whether sensitive locations are accessible from unprivileged signal sources. Further research is required in explicitly modelling and verifying non-register states and timing flows. 
Potential research directions include exploring hybrid approaches where formal methods can be used to guide optimized coverage (via fuzzing) of dynamic testing of RTL.

\section{Related Work}\label{sec:related}
We present here related work in hardware security verification while identifying their limitations with respect to detecting HardFails. 
We also provide an overview of recent software attacks exploiting the underlying hardware vulnerabilities.

\subsection{Current Detection Approaches}
Security-aware design and verification of hardware have gained significance and traction only recently as the critical security threat posed by hardware vulnerabilities became acutely established. Confidentiality and integrity are the most commonly investigated properties~\cite{survey16onur} in hardware security.
They are usually expressed using information flow properties between entities at different security levels.
Besides manual inspection and simulation-based techniques, systematic approaches proposed for verifying security properties in hardware include: formal verification methods such as proof assistance, model-checking, symbolic execution, and information flow tracking.

\textbf{Proof assistant and theorem-proving} methods rely on mathematically modeling the system and the required security properties into logical theorems and formally proving if the model complies with the properties. VeriCoq~\cite{vericoq15} based on the Coq proof assistant transforms the Verilog code that describes the hardware design into proof-carrying code.
VeriCoq supports the automated conversion of only a subset of Verilog code into Coq and subsequent works~\cite{vericoq17part1, vericoq17part2} automate the creation of the theorems and proofs and check information flow properties. However, this assumes accurate labeling of the initial sensitivity labels of each and every signal in order to effectively track the flow of information. This is cumbersome, error-prone and would never scale in practice beyond toy examples to complex real-world designs. Timing (and other) side-channel information flows are not modeled and there is room for false positives. Finally, computational scalability to verifying real-world complex SoCs remains an issue given that the proof verification for a single AES core requires ~30  minutes to complete~\cite{vericoq17part2}.

\textbf{Model checking}-based approaches are widely used in industry-standard tools. A given property is checked against the modeled state space and possible state transitions using provided invariants and predefined conditions. Such techniques remain limited in terms of scalability as computation time scales exponentially with the model and state space size. This can be alleviated by using abstraction to simplify the model or constraining the state space to a bounded number of states using assumptions and conditions.
However, this introduces false positives and missed vulnerabilities and requires expert knowledge.
Most industry-leading tools, such as the one we use in this work, rely on model checking algorithms such as boolean satisfiability problem (SAT) solvers and property specification schemes, e.g., assertion-based verification to verify the required properties of a given hardware design. 

\textbf{Side-channel leakage modeling and detection} remain an open problem. Recent work~\cite{leakage14ruby} uses the Mur$\varphi$ model checker to verify different hardware cache architectures for side-channel leakage against different adversary models. A formal verification methodology for SGX and Sanctum enclaves under a limited adversary was introduced in~\cite{sanctum17formal}. 
However, such approaches are not directly applicable to the hardware implementation. They also rely exclusively on formal verification and remain inherently limited by the underlying algorithms in terms of scalability and state space explosion, besides demanding particular effort and expertise to use.

\textbf{Information flow analysis} (such as SPV) are better suited for this purpose where a data variable or input is assigned a security label (or a \emph{taint}), and the taint propagation is monitored. This way, the designer can verify whether the system adheres to the required security policies. Recent works have demonstrated the effectiveness of hardware information flow tracking (IFT) in identifying security vulnerabilities, including unintentional timing side channels and intentional information leakage through hardware Trojans.
IFT techniques are proposed at different levels of abstraction: gate-, RT, and language-levels.  
Gate-level information flow tracking (GLIFT)~\cite{glift09, glift10, glift11} performs the IFT analysis directly at gate-level by generating GLIFT analysis logic that is derived from the original logic  and operates in parallel to it. Initially~\cite{glift09, glift10}, the GLIFT logic was fabricated along with the original logic, hence, incurring unreasonably high overheads in area, power, and performance. 
More recent works~\cite{glift11,glift11usb} applied GLIFT to the gate netlist only during simulation/verification and stripped it off before fabrication.
While gate-level IFT logic is easier to automatically generate, it does not scale well with design size.
Furthermore, the authors in~\cite{glift16} reason that when information flow tracking uses strict non-interference, this taints any information flow as a vulnerability. However, in reality, this is not the case. By relaxing the strict property of non-interference, "how much" of the tainted data flows is quantified using information theoretic methods and a GLIFT/information theoretic joint analysis technique is proposed. However, this requires extensive statistical analysis which is also not scalable with complex hardware designs.
The aforementioned IFT techniques track the information flow conservatively. The label of any operation output is assigned according to the ``highest'' security label of any of its inputs, irrelevant of functionality. While this over-approximation increases scalability for more complex hardware, it is imprecise and results in too many false positives. 

RTL-level IFT was proposed in~\cite{rtlift17} where the IFT logic is derived at a higher abstraction level, is faster to verify, and the accuracy vs. scalability trade-off is configurable.  In principle, at RTL-level all logic information flows can be tracked, and in~\cite{rtlift17} the designer is allowed to configure the complexity (whether to track explicit and implicit information flows) and precision of the tracking logic. 
Another approach isolates timing flows from functional flows and shows how to identify timing information leakage for arithmetic and cryptographic units~\cite{clepsydra17}. 
However, whether it can scale well to effectively capture timing leakage in real-world complex processor designs remains an open question. 

At the language level, Caisson~\cite{caisson11} and Sapper~\cite{sapper14} are security-aware HDLs that use a typing system where the designer assigns security ``labels'' to each variable (wire or register) by the security policies required. However, they both require redesigning the RTL using a new hardware description language which is not practical. 
SecVerilog~~\cite{secverilog15, secverilog17} overcomes this by extending the Verilog language with a dynamic security type system. Here, designers assign a security label to each variable (wire or register) in the RTL Verilog code to enable a compile-time check of hardware information flow. However, it must use predicate analysis during simulation to reason about the run-time behavior of the hardware state and dependent data types for precise flow tracking.

\textbf{Hardware/firmware co-verification} to capture and verify hardware/firmware interactions remains an open challenging problem and is not available in widely used industry-standard tools. 
A co-verification methodology~\cite{Huang18} addresses the semantic gap between hardware and firmware by modeling hardware and firmware using instruction-level abstraction to leverage software verification techniques. However, this requires modeling the hardware that interacts with firmware into an abstraction which is semi-automatic, cumbersome, and lossy.
While research is underway~\cite{tortuga-unison} to analyze a limited amount of low-level firmware running on top of a simulated RTL design these approaches are still under development and not yet widely available.
Finally, current verification approaches focus on register-state based information-flow analysis, e.g., to monitor whether sensitive locations are accessible from unprivileged signal sources, and further research is required to explicitly model non-register states and timing explicitly alongside the existing capabilities of those tools.

\subsection{Recent Attacks}
\begin{table*}[t]
\small
	\renewcommand{\arraystretch}{1.25}
	\centering
	\begin{tabular}{@{}l c c c c c c c c@{}}
		\toprule
		\multirow{2}{1cm}{Attack} & \multirow{2}{2cm}{Privilege Level} & \multirow{2}{1.25cm}{Memory Corruption} & \multirow{2}{1.5cm}{Information Leakage} & \multirow{2}{1.5cm}{Cross-modular} & \multirow{2}{1.25cm}{HW/FW-\\Interaction} & \multirow{2}{2cm}{Cache-State Gap} & \multirow{2}{1.75cm}{Timing-Flow Gap} & \multirow{2}{1.25cm}{\textbf{HardFail}} \\
        \\
		\midrule

		Cachebleed~\cite{cachebleed2017yarom} & unprivileged & \xmark & \cmark  & 
		\xmark & \xmark & \xmark & \cmark &\cmark \\ 
		
		TLBleed~\cite{gras2018translation} & unprivileged & \xmark & \cmark  &
        \cmark & \xmark & \cmark & \cmark & \cmark \\
        
        BranchScope~\cite{evtyushkin2018branchscope} & unprivileged & \xmark & \cmark  &
        \xmark & \xmark & \cmark & \xmark &\cmark \\ 		

		Spectre~\cite{spectre}  & unprivileged & \xmark & \cmark &
		\cmark & \xmark & \cmark & \xmark &\cmark \\ 

		Meltdown~\cite{meltdown2018lipp}  & unprivileged & \xmark & \cmark   & 
		\cmark & \xmark & \cmark & \xmark &\cmark \\ 
		
		MemJam~\cite{memjam2018moghimi} & supervisor & \xmark & \cmark  &
		\cmark & \xmark & \xmark & \cmark &\cmark \\  

		CLKScrew~\cite{clkscrew2017tang} & supervisor & \cmark & \cmark  & 
		\xmark & \cmark & \xmark & \cmark &\cmark \\

		Foreshadow~\cite{foreshadow}  & supervisor & \cmark & \cmark   & 
		\cmark & \cmark & \cmark & \xmark &\cmark \\
		\bottomrule
                \\
	\end{tabular}
	\caption{\textbf{Classification of existing vulnerabilities:} when reviewing recent microarchitectural attacks with respect to existing hardware verification approaches, we observe that the underlying bugs would have been difficult to detect due to their HardFail properties which we infer from the technical descriptions and errata of these recently published hardware vulnerabilities.}
	\label{tab:attacks}
\end{table*}

As outlined in Section~\ref{sec:discussion}, some recent attacks combine different problems (e.g., inherent cache leakage and implementation errors).
We explain and classify the underlying hardware vulnerabilities (see Table~\ref{tab:attacks}), as inferred from the technical description of these exploits.

Yarom et al. demonstrate that software-visible side channels can exist even below cache-line granularity in their CacheBleed~\cite{cachebleed2017yarom} attack---undermining a core assumption of prior defenses such as scatter-gather~\cite{brickell2006software}.
We categorize it as a timing-flow bug, since software can cause clock cycle differences for accesses mapping to the same bank below cache line granularity to break (assumed) constant-time implementations.

The recent TLBleed~\cite{gras2018translation} attack demonstrates that current TLB implementations can be abused to break state-of-the-art cache side-channel protections.
As outlined in Section~\ref{sec:ariane}, TLBs are typically highly interconnected with complex processor modules such as the cache controller and memory management unit,
making vulnerabilities therein very hard to detect through automated verification or manual inspection.

BranchScope~\cite{evtyushkin2018branchscope} extracts information through the directional branch predictor. Hence, it is unaffected by software mitigations that prevent leakage via the BTB.
The authors also propose alternative design strategies for the BTBs, such as randomizing the patter-history table.
We classify it as a cache-state gap in branch prediction units, which is significantly challenging to detect using automated RTL security verification techniques, since existing tools have a limited view of non-register states.
In the Meltdown attack~\cite{meltdown2018lipp},  speculative execution can be exploited on modern processors (affecting all main vendors) to completely bypass all memory access restrictions. Van Bulck et al.~\cite{foreshadow} demonstrated how to apply this to popular processor-based security extensions.
Modern out-of-order processors optimize utilization of idle execution units on a CPU core, e.g., by speculatively scheduling pipelined instructions ahead of time.
Kocher et al.~\cite{spectre} show that this can be exploited across different processes in a related attack, as arbitrary instruction executions would continue during speculation.
While these accesses are later correctly rolled-back (i.e., they are not \emph{committed} to the final instruction stream) their effect on the caches remains visible to software in the form of a timing side channel.
This means that addresses that are accessed illegally during speculation will be subsequently be loaded \emph{faster}, since they are cached.
The authors present end-to-end attacks, e.g., to leak arbitrary physical memory on recent platforms, regardless of the operating system or system configuration.
We classify these vulnerabilities as hard to detect mainly due to scalability challenges in existing tools, since the out-of-order scheduling module is connected to many subsystems in the CPU to optimize utilization.

MemJam~\cite{memjam2018moghimi} exploits false read-after-write dependencies in the CPU to maliciously slow down victim accesses to memory blocks within
a cache line.
Similar to Cachebleed, this breaks any constant-time implementations that rely on cache-line granularity, and we categorize the underlying vulnerability as being hard to detect in existing RTL implementations due to timing-flow gap and many cross-module connections.

CLKScrew~\cite{clkscrew2017tang} abuses low-level power-management functionality that is exposed to software on many ARM-based devices, e.g., to optimize battery life.
Tang et al. demonstrated that this can be exploited by malicious users to induce faults and glitches dynamically at runtime in the processor.
By maliciously tweaking clock frequency and voltage parameters, they were able to make the execution of individual instructions fail with a high probability.
The authors constructed an end-to-end attack that works completely from software and breaks the TrustZone isolation boundary, e.g., to access secure-world memory from the normal world.
We categorize CLKScrew to have vulnerable hardware-firmware interactions and timing-flow gap, since it directly exposes clock-tuning functionality to attacker-controlled software.

\section{Conclusion}\label{sec:conclusion}

Software security bugs and their impact have been known for many decades now with a spectrum of established techniques to detect and mitigate them. However, hardware security bugs only became recently significant with the growing complexity of modern processors and their effects have been shown to be even more detrimental than that of software bugs. Moreover, the techniques and tools to detect them are still at their infancy.  While some hardware bugs can be patched with microcode updates, many are not. As demonstrated by our results, many hardware bugs go undetected by manual inspection and verification techniques---even by using industry-standard tools and crowds-sourced expertise. The security impact of some of these bugs is further exacerbated if they are software-exploitable. 

In this paper, we have identified a non-exhaustive list of properties that make hardware security bugs difficult to detect:  complex cross-module inter-dependencies, timing channel leakages, subtle cache states, and hardware-firmware interactions. While these effects are common in modern SoC designs, they are difficult to model, capture and verify using both manual inspection and verification techniques. 
Our investigative work highlights why we must treat the detection of hardware bugs as significantly as that of software bugs. Through our work, we urge further research to advance the state of the art in hardware security verification. Particularly, our results indicate the need for increased scalability, efficacy, accuracy and automation of these tools, making them easily applicable to large-scale industry-relevant SoC designs. 

{\footnotesize
\bibliographystyle{abbrv}
\bibliography{main}
}

\appendices
\section*{Appendix}
\renewcommand{\thesubsection}{\Alph{subsection}}

\subsection{Ariane Core and RTL Hierarchy}\label{sec:ariane_figs}
\begin{figure*}[p]
    \centering
    \includegraphics[width=.6\textwidth]{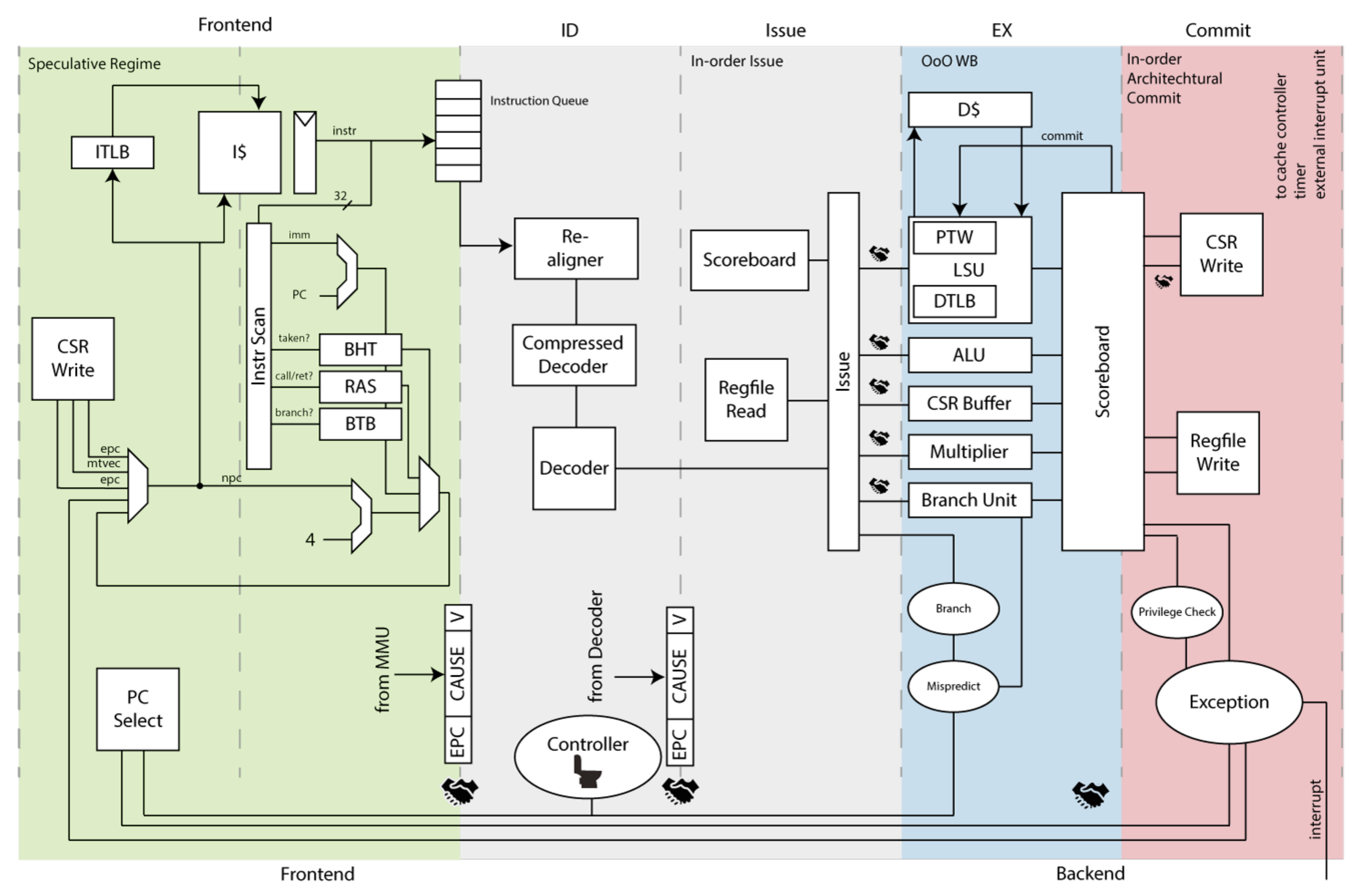}
    \caption{High-level architecture of the Ariane core~\cite{ariane}}
    \label{fig:ariane_core}
\end{figure*}

We present here supporting material for the Ariane core which we investigate in \ref{sec:ariane}.
Figure \ref{fig:ariane_core} shows the high-level microarchitectural layout of the Ariane core to visualize better its complexity.
This RISC-V open-source core pales in comparison with the complexity of a modern x86 or ARM processor and their far more sophisticated microarchitecture and optimization features.

\begin{figure*}[p]
    \centering
    \includegraphics[width=.9\textwidth]{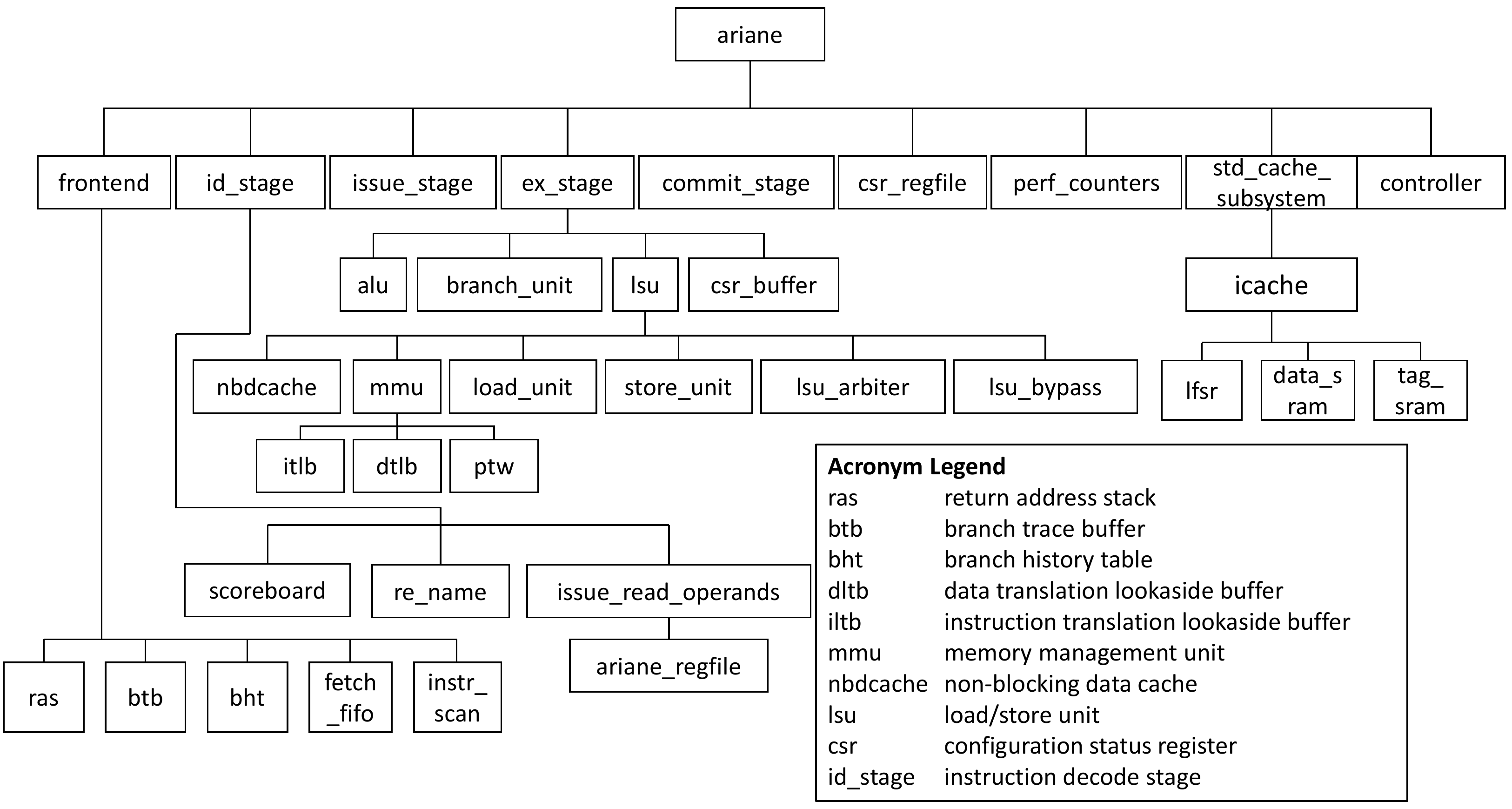}
    \caption{Illustration of the RTL module hierarchy of the Ariane core}
    \label{fig:ariane_hierarchy}
\end{figure*}

Figure \ref{fig:ariane_hierarchy} illustrates the hierarchy of the RTL components of the Ariane core. This focuses strictly on the core and excludes all uncore components, such as the AXI interconnect, peripherals, the debug module, boot ROM, and RAM.

\subsection{Details on the Pulpissimo Bugs}\label{sec:apdx_pulpissimo}

In this appendix we discuss both some of our implemented bugs and the native bugs that were already in the SoC and discovered by some of the competition teams.

\textbf{Bugs in crypto units and incorrect usage:} We extended the Soc with a faulty cryptographic hardware processing unit which had a selection multiplexer to select between AES, SHA1, MD5, and the temperature sensor. A hardware processing engine operates in parallel independent of the main processing core of the chip. The mux itself was modified so that a race condition occurs if more than one enable bit in the status register is enabled, causing unreliable behavior in these security critical modules.

Furthermore, SHA-1 is an outdated cryptographic hash function. SHA-1 has not been considered secure since 2005 and has been compromised by a number of attacks and replaced over the years with SHA-2 and SHA-3 instead. This type of bug is not detectable by formal verification and requires expert specification and design decisions and manual inspection. This is strictly a specification/design issue and not an implementation bug, therefore it is out of the scope of automated approaches and formal verification methods. These are as good as the provided specification and security properties and cannot infer the \emph{intended} security requirements, but only that the implementation matches the described security requirements. 

Finally the cryptographic key used by this unit is stored and read from unprotected memory, which allows for possible untrustworthy access and secret key. The fact that there is no dedicated temperature register, and instead the temperature sensor register is muxed with the different crypto modules which operate at a different security level is also a potential threat. The temperature sensor register value can muxed as output instead of the crypto engine output and vice versa, all of which are illegal information flows, which could compromise the cryptographic operations.
\begin{lstlisting}[language=Verilog, caption=\textbf{Incorrect use of crypto RTL:} The key input for the AES (g\_input) is connected to signal b. This signal is then passed through various modules until it connects directly to the L2 module.]
input  logic [127:0] b,
...
aes_1cc aes(
  .clk(0),
  .rst(1),
  .g_input(b),
  .e_input(a),
  .o(aes_out)
  );
\end{lstlisting}

\textbf{Bugs in security modes:} We have replaced the standard PULP\_SECURE parameter in the riscv\_cs\_registers and riscv\_int\_controller files with a parameter named PULP\_SEC which is always rigged at logical level "1", effectively disabling the secure mode checks for these two modules. Another security bug we have inserted is switching the write and read protections for the AXI bus interface, which causes erroneous protection checks for read and writes.

\textbf{Bugs in the JTAG module:} We have also implemented a JTAG password-checker and injected a multitude of bugs in it, including the password being hardcoded in the password checking file itself. The password checker also only checks the first 31 bits, which reduces the effort needed by an attacker to brute force the password. The password checker also does not reset the state of the correctness of the password when an incorrect bit is detected, allowing for repeated partial checks of passwords to end up unlocking the password checker. This is also facilitated by the fact that the index overflows after you hit bit 31, allowing for an infinite cycling of bit checks.

\end{document}